\documentclass[11pt, letterpaper]{article} 

\usepackage[includeheadfoot,
            marginratio={1:1,2:3}, 
            width=440pt, 
            height=688pt,]{geometry}

\usepackage{amsmath}
\usepackage{amsfonts}
\usepackage{amssymb}
\usepackage{bbm,bm}
\usepackage{mathtools}
\usepackage{slashed}
\usepackage{mathalfa}

\usepackage{graphicx,caption,subfigure}
\usepackage{tikz} 
\usepackage{tikz-cd}
\usetikzlibrary{decorations.pathreplacing,calc,tikzmark}
\usepackage{booktabs}
\usepackage{adjustbox}
\usepackage[utf8]{inputenc}
\usepackage[splitrule]{footmisc} 
 
\usepackage{placeins}
\usepackage{empheq}
\usepackage{paralist}
\usepackage{cite}
\usepackage[normalem]{ulem}
\usepackage{soul}

\usepackage[colorlinks=false,urlbordercolor=red]{hyperref}
\usepackage{xcolor}
\usepackage{tensor}

\usepackage{lipsum}

\newcommand{\nc}{\newcommand}
\nc{\lb}{\llbracket}
\nc{\rb}{\rrbracket}
\nc{\gl}{\llbracket}
\nc{\gr}{\rrbracket}
\nc{\del}{\partial}
\nc{\eq}[1]{\begin{equation}
                     \begin{split} #1 \end{split}
                     \end{equation}
}
\nc{\ov}{\overline}
\nc{\fa}{\hat}
\nc{\fb}{\MakeUppercase}
\nc{\fc}{\tilde}
\nc{\myhash}{\raisebox{\depth}{\#}}

\allowdisplaybreaks[2]
\numberwithin{equation}{section}
\DeclareUnicodeCharacter{2212}{-}
\begin{document}


\vspace*{-1.5cm}
\begin{flushright}
  {\small
  MPP-2025-122
  }
\end{flushright}

\vspace{1.0cm}
\begin{center}
  {\huge  Emergence   of   CY Triple Intersection  
    \\[0.3cm]
    Numbers in M-theory  } 
\vspace{0.4cm}

\end{center}

\vspace{0.25cm}
\begin{center}
{
\Large Ralph Blumenhagen$^{1}$ and Aleksandar
Gligovic$^{1,2}$
}
\end{center}

\vspace{0.0cm}
\begin{center} 
 \emph{$^{1}$ 
Max-Planck-Institut f\"ur Physik (Werner-Heisenberg-Institut), \\ 
Boltzmannstra\ss e  8,  85748 Garching, Germany } 
\\[0.1cm] 
\vspace{0.25cm} 
\emph{$^{2}$ Exzellenzcluster ORIGINS, Boltzmannstr. 2, D-85748 Garching, Germany}\\[0.1cm]
\vspace{0.3cm}
\end{center} 
\vspace{0.5cm}

\begin{abstract}
  To give  more credence to the M-theoretic Emergence Proposal
  it is important to show that also classical kinetic terms in a
  low energy effective action arise as a quantum effect from integrating out light towers of states.
   We show   that for  compactifications
  of type IIA on Calabi-Yau manifolds,
  the classical weak coupling Yukawa couplings, which are
  the triple intersection numbers of the Calabi-Yau threefold, can be obtained
  from the  1/2-BPS protected one-loop Schwinger integral over $D2$-$D0$ bound states,
  after employing a novel  regularization for the final infinite sum
  of  Gopakumar-Vafa invariants.
  Approaching the problem in a consecutive manner from 
  6D decompactification
  over emergent string to the ultimate  M-theory limits, we
  arrive at a mathematically concrete regularization  that involves
  finite distance degeneration limits of Calabi-Yau threefolds  in an
  intriguing way. We test and challenge this proposal by
  the concrete determination of the periods around such
  degeneration points for threefolds
  with one K\"ahler modulus and the two examples $\mathbb
  P_{1,1,1,6,9}[18]$   and $\mathbb P_{1,1,2,2,6}[12]$.
\end{abstract}

\vfill
\rule{4cm}{0.5pt}

\vspace{-0.1cm}
blumenha@mpp.mpg.de, aglig@mpp.mpg.de

\thispagestyle{empty}
\clearpage

\setcounter{tocdepth}{2}

\tableofcontents


\section{Introduction}

The Emergence Proposal
\cite{Heidenreich:2017sim,Grimm:2018ohb,Heidenreich:2018kpg} is among
the most promising and yet less understood ideas of the swampland
program, aiming at characterizing universal features of gravitational
effective theories. Its strongest version \cite{Palti:2019pca} holds
that the dynamics (kinetic terms) for all fields are emergent in the
infrared by integrating out towers of states down from an ultraviolet
scale, which is below the Planck scale. The species scale
\cite{Veneziano:2001ah,Dvali:2007hz} provides a natural candidate  for such an
ultraviolet scale,  which has been employed
in the subsequent field theory approaches of \cite{Marchesano:2022axe,Castellano:2022bvr,Castellano:2023qhp, Casas:2024ttx}.

However, as long as one sticks to this strong version of the Emergence
Proposal, there seems to be little room to implement it exactly in
full string theory,
see e.g.~\cite{Ooguri:2006in,Lee:2021usk,Blumenhagen:2023yws,Blumenhagen:2023tev}.
In the spirit of the Emergent String Conjecture \cite{Lee:2019wij},
the only other candidate are decompactification limits.
And indeed, given our ignorance of M-theory and the quantum
origin of graviton scattering in the BFSS matrix model,
it is a logical possibility
that the (strong) Emergence Proposal might be realized
therein. This was first suggested in
\cite{Blumenhagen:2023tev} and then elaborated on in a series of papers
\cite{Blumenhagen:2023xmk,
  Blumenhagen:2024ydy,Blumenhagen:2024lmo,Artime:2025egu}. In particular, in
\cite{Blumenhagen:2024ydy} the precise form of the $R^4$ interaction
in M-theory, including the correct relative coefficients, was
recovered by integrating out at one-loop level infinite towers of states with typical
mass not larger than the eleven dimensional Planck scale, which is the
species scale in the studied limit and duality frame. This computation
was generalized to a higher derivative $F^4$ interaction
in type IIA compactified on a $K3$ to six dimensions \cite{Artime:2025egu}.

However, as it is clear from its initial formulation, the Emergence
Proposal is first and perhaps foremost about kinetic terms, so one
should be able to recover them by integrating out towers of
states. This is challenging in general, especially in a gravitational
context. For example, due to the equivalence principle, the
Einstein-Hilbert term shall in principle receive contributions from
all states in the theory, for gravity couples to everything. To
simplify the task, one can look at less realistic kinetic
interactions, possibly with some protection under a certain
symmetry. A good example of such an interaction can be found in
Calabi-Yau (CY) compactifications of the type IIA superstring leading to a
four-dimensional theory with eight preserved supercharges. Here,
kinetic terms are encoded into a holomorphic function of the K\"ahler
moduli, the prepotential. More specifically, the tree-level
contribution
at weak type IIA string coupling is encoded in a cubic polynomial
with coefficients related to the triple intersection numbers
(TINs) of the CY.
Hence, if the M-theoretic Emergence Proposal is to be correct,
then one should be able to reproduce all TINs
by integrating out the light infinite towers of states in a co-scaled strong coupling limit of type IIA, where one scales all Kähler moduli isotropically such that the 4D Planck scale remains fixed. Henceforth, we will refer to this limit as the (isotropic) M-theory limit. As generally shown in \cite{Blumenhagen:2023xmk}, the light towers with a mass scale
below the species scale, i.e.\ the eleven dimensional Planck-scale,
are transverse $M2$- and $M5$-branes carrying KK momentum
along the eleventh (type IIA $D0$-branes)  and any other compact dimension.

A first step in this
direction was performed in \cite{Blumenhagen:2023tev} for a
non-compact CY, namely the resolved conifold,
where the geometry only admits a single 2-cycle, on which
an  $M2$-brane could be wrapped.
After realizing that the M-theory amplitude is nothing
else than the Schwinger integral of Gopakumar/Vafa \cite{Gopakumar:1998ii,Gopakumar:1998jq},
which is precisely integrating out 1/2 BPS bound states of
$D2$- and $D0$-branes, a proper regularization
of its UV-divergence followed by $\zeta$-function regularization
also gave the correct formal TIN
of the resolved conifold.
It is the purpose of this paper to generalize this
emergence computation of TINs to compact CY threefolds,
where finally one  has to sum over the full homology lattice
$H_2(X,\mathbb Z)$ weighted by the exponentially growing
Gopakumar-Vafa (GV) invariants\footnote{Let us notice that an alternative approach has been put
forward in a series of recent papers by Hattab/Palti
\cite{Hattab:2023moj,Hattab:2024thi,Hattab:2024chf}. In particular,
the initial real Schwinger integral of Gopakumar/Vafa has been changed
to a complex contour integral.}.
It seems to be a horrendous
task to systematically define a regularization of these highly
divergent  infinite sums that is guaranteed to give the correct
TINs.

For appreciating the approach we take, it is important
to recall that in  the former  emergence computations
\cite{Blumenhagen:2023tev,Blumenhagen:2023xmk,
  Blumenhagen:2024ydy,Blumenhagen:2024lmo,Artime:2025egu},
the one-loop amplitude was initially defined
via an integral over a real Schwinger parameter, similar
to naive string amplitudes before implementing level matching
and modular invariance.
Such integrals are notoriously UV-divergent, but in
\cite{Blumenhagen:2023tev,Blumenhagen:2024ydy} a way to regularize them was proposed, which
involved a minimal substraction of the divergent terms
followed by a $\zeta$-function regularization of the final
infinite sums. It was shown in \cite{Blumenhagen:2024ydy} that this
procedure also gave the correct finite results upon
applying it to one-loop string amplitudes, which
constitutes a non-trivial consistency check.
The question is whether a generalization of this
regularization  can be found that produces
finite TINs for infinite sums of the type
\eq{
  \label{introeq}
   Y^{(0)}_{t_i t_ j t_k}= {1\over 2}\sum_{\beta\in H_2(X,\mathbb Z)} \alpha_0^\beta \,\beta_i   \beta_j   \beta_k   \,,   
}
where $\alpha_0^\beta$ denote the Gopakumar-Vafa invariants
which are known to generically grow as $\alpha_0^\beta\sim
\exp(\gamma \beta)$ with $\gamma>0$ \cite{Bershadsky:1993cx}.

This paper is organized as follows: In section 2 we explain
in more detail the origin of these infinite sums
and provide a mathematical  toy example, which already contains
an essential hint of how one could approach this problem.
Before directly dealing with the most involved isotropic M-theory
limits, in section 3 we first consider a couple of special CYs, which
either admit an elliptic or a $K3$-fibration. For the elliptic
fibrations we take a 6D decompactification limit so that only those
$M2$-branes wrapping the elliptic fiber are the light towers of
states contributing to the Schwinger integral. Integrating them out
can be performed just via $\zeta$-function regularization
and precisely yields the correct constant to complete
the modular form $E_4(t)$, where $t$ measures the size of the fiber. This serves as a first hint that indeed the regularization of infinite sums of the type \eqref{introeq} leads to sensible results.
For the $K3$-fibration we take the emergent string limit which
admits a dual weakly coupled heterotic string description.
In this case one expects that only the triple intersection
numbers not involving the base 2-cycle are emerging from
integrating out $D2$-$D0$ bound states. Employing
the heterotic dual we can determine the relevant GV
invariants and concretely formulate and test a proposal of
how the regularization of \eqref{introeq} can be performed.
This involves a  $t\to 0$ limit and modular S-transformation
in an intriguing way.

Building on these working examples,
in section 4 we approach the actual problem of the emergence
of the complete set of TINs in isotropic M-theory limits.
As will become clear this part is more speculative,
as the aforementioned limit $t\to 0$ cannot be taken.
This is first discussed for the Quintic threefold, where
the prepotential already diverges at the location of the
conifold singularity, which is correlated with the exponential
growth of the GV
invariants. The analysis of this example will guide us to a mathematically
concrete  regularization
of the infinite sum \eqref{introeq} that involves
the wanted TIN and  the limit 
of the known prepotential in the type
IIA weak coupling regime  towards degeneration limits of the CY.
For the Quintic and the  other 13  CY threefolds from \cite{Bonisch:2022mgw} with $h_{11}=1$
this gives the correct TIN from the limit 
to the  location of the conifold singularity.

In the rest of the paper
we  challenge  this admittedly speculative proposal for two
CYs having  two K\"ahler moduli, namely the elliptic fibration
$\mathbb P_{1,1,1,6,9}[18]$ featuring two intersecting conifold loci
and the $K3$-fibration $\mathbb P_{1,1,2,2,6}[12]$ having two intersection
points of a conifold and another degeneration locus, often called
(heterotic dual) strong coupling singularity.
We will find that for conifold loci our proposal works quite
straightforwardly while  the other degeneration locus
reveals some subleties that require a further refinement
of the regularization procedure.

On the more technical side, these computations involve the knowledge
of the CY periods in the vicinity of these degeneration loci in the
complex structure moduli space\footnote{We are indebted to Rafael \'Alvarez-Garc\'ia for sharing his insights into these computations and for actually providing some of the results and teaching us how to
achieve them.}, i.e.\ results that are not generally  available in the literature, yet. The details of these computations will be reported in \cite{unpublishedRafa}. We also add
two appendices. In the first, we review the relation
between the asymptotic growth of GV invariants and
the appearance of  degeneration loci in the complex
structure moduli space and also mention a few
observations once one tries to generalize
the original computation for the Quintic  \cite{Candelas:1990rm}
to the CY $\mathbb P_{1,1,2,2,6}[12]$.
In the second appendix we point out that  
in case of a description of the CY  in terms of a Gauged Linear
Sigma Model (GLSM), the location of the singularity is nothing else
than the  quantum corrected singular point in the middle of the classical phase diagram of this GLSM.

\section{Preliminaries}

In \cite{Blumenhagen:2023tev} it has been suggested that the strong Emergence Proposal might be realized in M-theory by looking at compactifications of type IIA string theory on a CY threefold $X$ yielding $N=2$ supergravity in four dimensions. 
The corresponding vector-multiplet moduli space is spanned by real
scalars $\tau_i$ which, together with the Kalb-Ramond axions $b_i$,
define the complex Kähler moduli $t_i = b_i + i\tau_i$. 
The kinetic terms and gauge couplings of the vector-multiplets are
determined by a holomorphic prepotential $\mathcal{F}_0(t)$. Due to
supersymmetry, this is tree-level exact in type IIA and reads\footnote{In this work we will omit writing the linear and quadratic terms, whose prefactors can be altered by acting with symplectic transformations \cite{deWit:1992wf}. However, part of that information is important for fixing an integral symplectic basis for the periods \cite{Hosono:1993qy,Hosono:1994ax}.}
\begin{equation}
\label{prepot}
\mathcal{F}_0(t) =  \frac{(2\pi i)^3}{g_s^2} \Big[
\frac{1}{6}\kappa^{ijk}\, t_i t_j t_k - \frac{\zeta(3)}{2(2\pi i)^3}
\chi(X) + {1\over (2\pi i)^3}\sum_{\beta \in H_2(X,\mathbb{Z})}
\alpha_0^{\beta} \, {\rm Li}_3(e^{2\pi i\beta \cdot t}) \Big] \,,
\end{equation}
where $\kappa^{ijk}=\kappa_{t_i t_j t_k}$ are the TINs of $X$,
$\chi(X)$ its Euler characteristic and $\alpha_0^{\beta}$ are genus
zero GV invariants \cite{Gopakumar:1998ii,Gopakumar:1998jq}.

It is the purpose of this paper to investigate whether one can  obtain
the classical cubic term in the prepotential from a one-loop
Schwinger-like integral in the strong coupling regime, i.e.\ the
isotropic M-theory limit. Following \cite{Blumenhagen:2023tev}, it is defined by  scaling all type IIA K\"ahler
moduli as $\tau_i\to \lambda \, \tau_i$ and $g_s\to \lambda^{3/2} g_s$
(with $\lambda\to\infty$) so that
the 4D Planck scale remains constant. Then, the $D0$-branes are the lightest species leading to the species scale  $\Lambda_{\rm sp}\sim M_{\rm
  pl}/\lambda^{1/2}$, which is at threshold with the mass of
$D2$-branes wrapping 2-cycles of the CY. This is a decompactification
limit to 5D described by  M-theory on the CY. When working in M-theory units,
this co-scaled limit leaves constant all radii of the CY measured in units
of the 11D Planck scale $M_*$ and scales
$r_{11}\to \lambda \, r_{11}$, $M_*\to M_*/\lambda^{1/2}$.
Then, the species scale is nothing
else than the 5D Planck scale, scaling in the same way as $M_*$.

Integrating out the light towers of states with mass scale below
or at the species scale we arrive at the prescription of Gopakumar/Vafa
\cite{Gopakumar:1998ii,Gopakumar:1998jq}, i.e.\ we need to integrate out wrapped $M2$-branes transverse with respect to the
M-theory circle but with KK-momentum along it. In type IIA variables,
these are $D2$-$D0$ bound states. 
The corresponding Schwinger integral reads
\begin{equation}
\label{F0_Schwinger}
\mathcal{F}_0(t) = \sum_{(\beta,n) \neq (0,0)} \alpha_0^{\beta}
\int_0^{\infty} \frac{ds}{s^3} \; e^{s Z_n(\beta)} \,,
\end{equation}
where $Z_n(\beta) = \frac{2\pi i}{g_s}(\beta \cdot t - n)$ is the
central charge of the supersymmetry algebra. 
The complete (perturbative) $\mathcal{F}_0$ for the (non-compact) resolved conifold,
which has only a single shrinkable 2-cycle and therefore no sum over
$\beta$, has been successfully reproduced in
\cite{Blumenhagen:2023tev}. There, the divergent sum over the number
of $D0$-branes was regularized with the analytic continuation of the
Riemann $\zeta$-function. For a general compact CY, the $D2$-$D0$
bound states wrap all 2-cycles in $H_2(X,\mathbb{Z})$ and therefore
another infinite sum needs to be performed.

\subsection{Quest for emergence of kinetic terms}

To set the stage, let us focus for the moment on compact threefolds
with $h_{11} = 1$ and a single TIN $\kappa$.
We can extract information about the cubic term from
\eqref{F0_Schwinger} by taking the third derivative with respect to
$t$, giving the ``Yukawa coupling''
\begin{equation}
Y_{ttt}=\frac{g_s^2}{(2\pi i)^3}\partial_t^3 \mathcal{F}_0(t) =
\frac{1}{g_s} \sum_{(\beta,n) \neq (0,0)} \alpha_0^{\beta} \beta^3
\int_0^{\infty} ds \, e^{\frac{2\pi i}{g_s} s(\beta t -n)} \,.
\end{equation}
After some steps, we can rewrite it as 
\begin{equation}
\label{derivF0}
 Y_{ttt}=
\sum_{\beta \neq 0} \alpha_0^{\beta}  \,\beta^3 \left(\frac{1}{2} +
  \frac{e^{2\pi i \beta t}}{1-e^{2\pi i \beta t} } \right) \,,
 \end{equation}
where the second term in the brackets is related to the
non-perturbative terms in \eqref{prepot}, i.e.~the $ {\rm
  Li}_3$-terms. Note that the factor $1/2$ is precisely the formal triple
intersection number for the single two-cycle of the resolved conifold
\cite{Gopakumar:1998ki}(see also \cite{Blumenhagen:2023tev}).
The whole expression, including the factor 1/2, is reminiscent of the thermal energy-density  of a gas of particles in four dimensions with energy levels $\beta$, degeneracy of states  $\alpha_0^{\beta} \beta^2$ and temperature $T=i/t$. 
The first term in the brackets is then the (divergent) zero point
energy, which in the following we call the  {\it zero point Yukawa
coupling} denoted as $Y_{ttt}^{(0)}$. This object is the focus of our interest.

Typically the (genus-zero) GV invariants scale exponentially for large
$\beta$, $\alpha_0^{\beta}\sim \exp(\gamma\beta)$, with $\gamma>0$. However, for $h_{11}>1$ there can also be
sub-towers of wrapped $D2$-branes leading to a weaker scaling such as $\alpha_0^{\beta}\sim\exp(\gamma\sqrt{\beta})$, or
even to a polynomial behaviour as for KK towers.
One can then take a corresponding infinite distance limit so that only
these towers contribute. The degeneracy of $\alpha_0^\beta$ is then
closely tied to the type of limit as shown in table~\ref{table_towers},
where we listed the limits relevant for  the following discussion.
\begin{table}[h] 
\renewcommand{\arraystretch}{1.2} 
\begin{center} 
\begin{tabular}{|c|c|} 
\hline
limit  &   degeneracy of $\alpha_0^\beta$       \\
\hline \hline
M-theory  &   $\exp(\gamma\beta)$ \\
emergent string  &   $\exp(\gamma\sqrt{\beta})$ \\
6D decompactification & $1$ \\
\hline
\end{tabular}
\caption{Infinite distance limits and degeneracy of genus-zero GV
  invariants.}
 \label{table_towers}
\end{center} 
\end{table}
In any case, the first sum in \eqref{derivF0} will be divergent
regardless of these limits. Hence, emergence means that upon
regularization, the $t$-independent first term in the bracket in
\eqref{derivF0} should give rise to the self-intersection number
\begin{equation}
\label{kappasum}
Y_{ttt}^{(0)}:= \frac{1}{2} \sum_{\beta = 1}^{\infty} \alpha_0^{\beta} \,\beta^3 \Big\vert_{\rm reg} \stackrel{!}{=}\kappa_{ttt} \,.
\end{equation}
Note that here the left-hand side is a quantity to be computed in the
strong coupling regime, $g_s\gg 1$, i.e.~within M-theory, whereas the
right hand side is the usual integer-valued tree-level Yukawa coupling
arising at weak string coupling, $g_s\ll 1$.
  
Therefore, the quest is to {\it define} a regularization such that the zero-point Yukawa coupling, $Y^{(0)}_{ttt}$, matches with the classical contribution in the perturbative string regime.
Recall that $Y^{(0)}_{ttt}$ diverges due to the sum over wrapped $D2$-$D0$ bound states in the M-theory regime where, in contrast to string perturbation theory, we currently have no formalism to perform the integral and get to a finite result right away. 
Indeed, such a computation shall be part of the microscopic description of M-theory, which is lacking at present. 
Therefore, here we can only head for a properly working regularization
scheme, like that of minimal subtraction and $\zeta$-function
regularization proposed for the computation of the $R^4$ terms in M-theory in \cite{Blumenhagen:2024ydy}.

For towers of $D2$-branes with a polynomial scaling of the GV
invariants, we can  still apply $\zeta$-function regularization, but for
the exponentially degenerate towers other methods need to be developed.
Our approach  is to move from simpler to more complicated cases and in
the process refine and adapt the regularization method. 
Let us start by  presenting  a regularization method for infinite sums
that already demonstrates our main strategy.

\subsection{Regularization via modular forms}
\label{sec_regmod}

Suppose we are dealing with sums of the form
\begin{equation}
    S_n = \sum_{k=1}^{\infty} k^n \, ,
\end{equation}
which are usually regularized with the $\zeta$-function $\zeta(-n)$.
It is instructive to recall how this works in practice. One introduces
a regulator $q = \exp(-2\pi/\Lambda)$, with $\Lambda \gg 1$, such that
for finite $\Lambda$ the sum is finite. Then, one takes the limit
$\Lambda\to\infty$ to isolate the divergence.
In practice, the conventional way is to introduce $q$ such that
derivatives of the geometric series arise, namely
\begin{equation}
S_n = \lim_{\Lambda \rightarrow \infty} \sum_{k=1}^{\infty} k^n q^k = \lim_{\Lambda \rightarrow \infty} \left[ (q \,\del_q)^n \left( \frac{q}{1-q} \right) \right] \,,
\end{equation}
for $|q|<1$.
One can then expand the right-hand side for large $\Lambda$ to obtain 
\begin{equation}
S_n =  \lim_{\Lambda \rightarrow \infty} \left[ n! \frac{\Lambda^{n+1}}{(2\pi)^{n+1}} + C_n  + \mathcal{O} (\Lambda^{-1})\right] .
\end{equation}
The divergence appears only at $(n+1)$-th order and can be minimally
subtracted, leaving us with the regularized value $C_n\equiv \zeta(-n)$.

Let us now arrive at the same result with an alternative method that
exploits properties of modular forms to extract the singular term. We
assume $n$ to be odd in what follows. The idea is to express the above
sum in terms of an Eisenstein series, which requires us to introduce
the regulator $q$ in a different fashion, namely 
\begin{equation}
S_{2n-1} = -2 \lim_{\Lambda \rightarrow \infty} \sum_{k=1}^{\infty} \sum_{l=1}^{\infty} k^{2n-1}\, q^{lk} \,.
\end{equation}
The normalization $(-2)$ is chosen in order to compensate with the
factor $\zeta(0)=-1/2$ arising from the sum over $l$ in the limit $q
\rightarrow 1$. We define the summation index $m = lk$ and then
replace the second sum with a sum over the divisors of $m$, giving  
\begin{equation}
S_{2n-1} = -2 \lim_{\Lambda \rightarrow \infty} \sum_{m=1}^{\infty} \sigma_{2n-1}(m) \, q^m \qquad \text{with} \quad \sigma_n(m) = \sum_{k | m} k^n \,.
\end{equation}
This form of $S_{2n-1}$ can now be directly related to the Eisenstein series
\begin{equation}
E_{2n}(\tau) =  1 + c_{2n} \sum_{m=1}^{\infty} \sigma_{2n-1}(m) \,  q^m \,, \quad
\end{equation}
with
\begin{equation}
c_{2n} = \frac{(2\pi i)^{2n}}{(2n-1)! \,\zeta(2n)} = \frac{2}{\zeta(1-2n)}\,.
\end{equation}
In this way, we arrive at
\begin{equation}
\label{sum_Eisenstein}
S_{2n-1} = \lim_{\Lambda \rightarrow \infty} \left[ - \frac{2}{c_{2n}} \left(E_{2n}\left({\frac{i}{\Lambda}}\right) -1 \right)  \right] \,.
\end{equation}
These Eisenstein series are modular forms of degree $2n$, so that under
a modular $S$-transformation they transforms as
\begin{equation}
E_{2n} \left(-\frac{1}{\tau} \right) = \tau^{2n} E_{2n}(\tau) \,.
\end{equation}
By applying this transformation rule to the first term in \eqref{sum_Eisenstein}, we find the behaviour
\begin{equation}
E_{2n}\left( \frac{i}{\Lambda}\right)=(-1)^n \Lambda^{2n}+ \mathcal{O}(e^{-2 \pi \Lambda})\,.
\end{equation}
A couple of comments are in order here. 
First, the singularity that has just been isolated has the same degree
as the one occurring in the former example with a slightly different
regularization scheme. 
Second, after  minimally subtracting this divergence, the
leading term is already exponentially suppressed in $\Lambda$ and
from the Eisenstein series one does not get any constant contribution
surviving the limit $\Lambda\to\infty$.
Hence, the only such contribution comes from the second term in
\eqref{sum_Eisenstein} and gives the regularized value
\begin{equation}
S_{2n-1}= \frac{2}{c_{2n}}= \zeta(1-2n)\,.
\end{equation} 
For our purposes, the advantage of this regularization scheme is that
it allows to explicitly isolate the divergence from the constant term
and from the contributions vanishing in the limit $\Lambda\to\infty$.
We will see that this method, and a generalization thereof,
can be successfully applied to regularize the sum over GV invariants as in \eqref{kappasum}.

\section{Large base limits  of fibered CY threefolds}

Before we discuss the isotropic M-theory limit, where the full homology lattice
contributes to the Schwinger integral, we look at simpler cases
where only part of it contributes.
This restriction is of course only allowed  if it is
correlated with taking an appropriate infinite
distance limit, in which the left out GV invariants
give rise to heavy towers of $D2$-branes with mass scale
above the species scale.
Even though for such limits one will not have one-loop emergence
of the full prepotential,  integrating out the light towers of
$D2$-branes will provide part of it and one  can still approach
the problem of what the meaning of the zero point Yukawa coupling
is.

\subsection{Elliptically fibered CY}
\label{sec:exellCY}

First, we consider  CY threefolds  with a sublattice  of
GV invariants that essentially grow like KK towers (see e.g.\ the
examples in \cite{Rudelius:2023odg}).
For such cases,  we
want to identify  an infinite distance  limit such that all wrapped $D2$-brane states with
exponential degeneracy are among the heavy states that, due to our
philosophy of perturbative QG theories, are  not
integrated out. As we will see, this allows us to carry out the sum over the simple
homology lattice by just using $\zeta$-function regularization.

The CY manifolds that feature the desired property are elliptically
fibered. As the maybe simplest example,
we consider the CY threefold obtained by resolving the $\mathbb Z_3$
singularity of a degree 18 hypersurface in
$\mathbb{P}_{1,1,1,6,9}[18]$ with Hodge numbers $(h_{21},h_{11})=(272,2)$. The GLSM data specifying the resolved  threefold are given in table \ref{P11169data} and can be found in the appendix of \cite{Hosono:1993qy}.
\begin{table}[h!]
    \begin{center}
    \begin{tabular}{|c|c c c c c c |c|} 
    \hline
    & $z_1$ & $z_2$ & $z_3$ & $z_4$ & $z_5$ & $z_6$ & $p$ \\ 
    \hline
    $l^{(1)}$ & 0 & 0 & 0 & 2 & 3 & 1 & -6 \\
    \hline 
    $l^{(2)}$ & 1 & 1 & 1 & 0 & 0 & -3 & 0 \\
    \hline
    \end{tabular}
    \caption{Data specifying elliptically fibered CY}
    \label{P11169data}
    \end{center}
  \end{table}

\noindent  
The threefold is an elliptic fibration with base $\mathbb{P}^2$,
i.e.\ $T^2 \rightarrow X \overset{\pi}{\rightarrow}
\mathbb{P}^2$. Coordinate divisors are generally
specified by conditions of the type $z_i = 0$. To be more concrete,
$z_6 = 0$ gives us $E = \mathbb{P}^2$, the base of the fibration, and
$z_1 = 0$ gives $L = \pi^*(l)$, where $l$ is a curve in the base and
$\pi^*(l)$ its respective pullback divisor.

Defining $H=3L+E$, the intersection numbers of $(H,L)$ were discussed in \cite{Candelas:1994hw} and read
\eq{
    H^3 = 9 \,, \quad H^2 \cdot L = 3 \,, \quad H \cdot L^2 = 1 \,, \quad L^3 = 0 \,.
}
The Poincaré dual two-forms are  denoted by $(w_H,w_L)$ so that we
expand  the Kähler form  as $J = \tau_1 w_H + \tau_2 w_L$. Hence, for the total volume we obtain
\eq{
    \mathcal{V}_6 \sim \int J^3 = 9\tau_1^3 + 9 \tau_1^2 \tau_2 + 3 \tau_1 \tau_2^2 \,.
}
Moreover, the fiber class $f\in H_2(X,\mathbb Z)$ is given by $L\cap
L$ and one can easily show that
\eq{
    {\cal V}_2(f)=\int_{L \cap L} J = \tau_1 \,.
  }
The second Mori-cone generator is given by the curve $l=L\cap E$ of volume
  \eq{
      {\cal V}_2(l)=\int_{ l = L \cap (H-3L)} J =\tau_2 \,.
   }

\subsubsection{Integrating out lightest towers of  states}

Our goal is to take an asymptotic limit where only those $D2$-branes
wrapping the elliptic fiber $f$ are considered as perturbative states,
since only these  GV invariants are constant in this particular case,
namely
\eq{
    \alpha_0^{(n_1,0)}=540 \qquad {\rm for}\quad n_1\ge 1\,,
  }
which is equal to minus the Euler characteristic of this CY. 
This can be achieved by the non-isotropic rescaling $\tau_2 \rightarrow
\lambda \tau_2$, $g_s\to \lambda g_s$  with $\lambda \rightarrow
\infty$ and $\tau_1$ fixed, so that 
$M_{{\rm pl},4}$ stays fixed. Hence a $D2$-brane wrapping an
effective  curve $C=n_1 f + n_2 l$ bound to $D0$-branes is 1/2 BPS and
has mass
\eq{
    m_{D0-D2} = \frac{M_s}{g_s} \big|n_1 t_1 + n_2 t_2 + m\big| = \frac{M_{{\rm pl},4}}{\sqrt{V_6}} \big|n_1 t_1 + n_2 t_2 + m\big|\,
  }
with $t_i=b_i+i\tau_i$. Since $V_6 \sim \lambda^2$, it is obvious that the lightest towers of
states
are $D0$-branes and $D2$-branes wrapping solely the class $f$,
i.e.\ having $n_2=0$ and mass
\eq{
                          m\sim  \frac{M_{{\rm pl},4}}{\lambda}\,.
}
Since $ \alpha_0^{(n_1,0)} = 540$ is constant, we are dealing with two light
multiplicative  KK towers \cite{Castellano:2021mmx}, which yield the species scale
\eq{
    \Lambda_{\rm sp} = M_{{\rm pl},4}^{\frac{1}{2}} \Delta m^{\frac{1}{2}} \sim \frac{M_{{\rm pl},4}}{\lambda^{1/2}} \,.
  }
In the classification of \cite{Lee:2019wij} this limit
translates into a $J$-Class A limit in the vector moduli space of
M-theory eventually leading to a further decompactification to
six dimensions (this F-theory like limit was previously called type III$_c$ in
\cite{Corvilain:2018lgw} and  also studied in
\cite{Marchesano:2023thx}). 
Let us mention that  among the light states there is also a light string, given by an $N\!S5$-brane wrapping the divisor $L$. One can show that there is no 4-cycle with a volume scaling like $\tau_1^2 \sim \lambda^0$ and the divisor volumes scale like
\eq{
   {\cal V}_4(H)= \int_H J \wedge J = (3\tau_1 + \tau_2)^2 \sim \lambda^2
   \,, \quad  {\cal V}_4(L)= \int_L J \wedge J = \tau_1(3\tau_1 + \tau_2) \sim \lambda \,.
}
For the tension of the $N\!S5$ wrapped on $L$ one finds $T_{NS5} =
\frac{M_s^2}{g_s^2}{\cal V}_4(L)$ and hence $M_{\rm str} =
\frac{M_s}{g_s} \sqrt{{\cal V}_4(L)} \sim \frac{M_{{\rm
      pl},4}}{\lambda^{1/2}}\sim \Lambda_{\rm sp}$.
As explained in \cite{Lee:2019wij}, this string is only a weakly coupled
critical string, if the threefold admits an additional $K3$ or $T^4$ fibration.
Therefore, we are not in the isotropic M-theory limit but in a limit where the
lightest towers
are bound states of $D0$- and $D2$-branes (wrapped  on $f$) signalling a
decompactification to 6D.

Using formula \eqref{derivF0},
we can now explicitly integrate out the light $D2$-$D0$ bound states
which gives
\eq{
  \label{ellipregulare}
    Y_{t_1 t_1 t_1} &= \sum_{k=1}^{\infty}
     \alpha_0^{(k,0)} \,k^3\left( {1\over 2} + {q^{k}\over (1-q^{k})} \right)
    = \frac{540}{2} \sum_{k = 1}^{\infty} k^3 +540 \sum_{k=1}^\infty   {k^3\,q^{k}\over (1-q^{k})}\\[0.1cm]
    &= {9\over 4} \left(1+240 \sum_{k=1}^\infty   {k^3\,q^{k}\over (1-q^{k})}\right)
    =  {9\over 4} E_4(t_1)\,,
  }
with $q=\exp(2\pi i  t_1)$ and where we used $ \sum_{k = 1}^{\infty} k^3=\zeta(-3)=1/120$.
From this calculation we do not obtain the triple-intersection number
$\kappa_{t_1 t_1 t_1}=9$, but since we are not taking the  M-theory limit with isotropic co-scaling of all compact directions, it was also not
expected a priori.
However, we see that the divergent  zero-point  Yukawa coupling, after $\zeta$-function
regularization provides precisely the constant term in $E_4(t_1)$ that
makes it a modular form of degree 4 (under modular transformation of $t_1$,
the size of the toroidal fiber). While the  appearance of a modular
form is expected for such an elliptic fibration, the actual point here
is to show how this result arises in the non-isotropic large coupling
limit from integrating out the light $D2$-$D0$ towers of states.
All this we take  as a first indication  that also for compact CYs
the regularized zero-point  Yukawa coupling  yields physically meaningful
results. Note that for getting this result, the generic factor $1/2$ in the first line of \eqref{ellipregulare} was essential. 

This computation is expected to generalize to every torus-fibered CY. One example is the CY $\mathbb P_{1,1,1,3,6}[12]$ with $(h_{21},h_{11}) = (165,3)$, which is a fibration over $\mathbb P^2$ with two sections. The GV invariants that count $D2$'s only wrapping the fiber are given by $\alpha_0^{(2k,0,0)}=324=-\chi(X)$ and $\alpha_0^{(2k-1,0,0)}=216$ with $k\ge 1$. For the relevant Yukawa coupling we obtain
\eq{
    Y_{t_1 t_1 t_1}&=\sum_{k\ge 1} k^3 \alpha_0^{(k,0,0)} \left({1\over 2}+ {q^k\over
        (1-q^k)}\right)\\
    &=
      216  \sum_{k\ge 1} k^3  \left( {1\over 2}+ {q^k\over
      (1-q^k)}\right) + 4\cdot 216 \sum_{k\ge 1} k^3  \left( {1\over 2}+ {q^{2k}\over
      (1-q^{2k})}\right)\\
  &={9\over 10} \Big( E_4(t_1)+ 4 E_4(2t_1) \Big)\,, 
}
which is a modular form of weight four for $\Gamma^+_0(2)$. The latter
is the extension of the congruence subgroup $\Gamma_0(2)$ by
the Fricke involution $\omega_2$, where $\omega_N$ (with $N \in \mathbb{N}$) is defined as 
\eq{\omega_N=\left(\begin{matrix} 0 & -{1\over \sqrt{N}}
         \\ \sqrt{N} & 0 \end{matrix}\right)\in PSL(2,\mathbb R) 
}
and induces the transformation $\tau\to -1/(N\tau)$ (see e.g.\ \cite{2022arXiv220606798I}). For torus fibrations with more sections, the cyclic properties of the GV invariants become more involved. For instance, the CY $\mathbb P_{1,1,1,3,3}[9]$ with $(h_{21},h_{11}) = (112,4)$ has three sections and the GV invariants with non-trivial fiber class are given by $\alpha_0^{(3k,0,0)}=216$ and $\alpha_0^{(3k-1,0,0)} = \alpha_0^{(3k-2,0,0)} = 162$ with $k\ge 1$. Nevertheless, the computation works similarly and yields 
\eq{
    Y_{t_1 t_1 t_1} = \frac{27}{40} \Big( E_4(t_1) + 9 E_4(3 t_1) \Big) \,,
}   
which is now a modular form of weight four for $\Gamma^+_0(3)$.

\subsection{\texorpdfstring{$K3$}{TEXT}-fibered CY}
\label{sec_k3fibered}

Our next example is a $K3$-fibration over the base $\mathbb{P}^1$. More specifically, we study the CY threefold obtained
from a hypersurface of degree 24 in $\mathbb{P}_{1,1,2,8,12}[24]$ with the
singularities
resolved. The model is known to be dual to the heterotic string on $K3
\times T^2$ in the limit where the base $\mathbb{P}^1$ grows large
\cite{Kachru:1995wm}. We will use this dictionary to extract  explicit
expressions for the Gopakumar-Vafa invariants of 2-cycles in the $K3$-fiber.
Concerning our conventions, we closely follow \cite{LopesCardoso:1996zu} (see also \cite{Curio:1997si}).

In table \eqref{P112812data} we provide the GLSM data for the
resolution $\mathbb{P}_{1,1,2,8,12}[24]$ from \cite{Hosono:1993qy}.
\begin{table}[ht]
    \begin{center}
    \begin{tabular}{|c|c c c c c c c |c|} 
    \hline
    & $z_1$ & $z_2$ & $z_3$ & $z_4$ & $z_5$ & $z_6$ & $z_7$ & $p$ \\ 
    \hline
    $l^{(1)}$ & 3 & 2 & 0 & 0 & 0 & 1 & 0 & $- 6$ \\
    \hline 
    $l^{(2)}$ & 0 & 0 & 0 & 1 & 1 & 0 & $-2$ & 0 \\
    \hline 
    $l^{(3)}$ & 0 & 0 & 1 & 0 & 0 & $-2$ & 1 &  0 \\
    \hline
    \end{tabular}
    \caption{Data specifying $K3$-fibered CY}
    \label{P112812data}
    \end{center}
  \end{table}
  
  \noindent
We denote the type IIA complexified Kähler moduli by $t_i$ $(i=1,2,3)$, where $t_2$
measures the size of the base.
In the limit of large base $t_2 \rightarrow \infty$ the prepotential
reads\footnote{Here we have set the prefactor $(2\pi i)^3/g_s^2=1$.}
\eq{
    \label{F0-IIA}
    \mathcal{F}_0^{\rm{IIA}} = \mathcal{F}_0^{\rm IIA,cubic} +
    \frac{1}{(2\pi i)^3} \sum_{n_1, n_3 = 0}^{\infty}
    \alpha_0^{(n_1,0,n_3)} \,{\rm Li}_3(q_1^{n_1} q_3^{n_3})
    +\mathcal{O} \left(e^{2\pi i  t_2}\right) \,,
  }
where world-sheet instantons wrapping also  the base become negligible.
The cubic term reads 
\eq{
    \label{F0-IIA-cubic}
    \mathcal{F}_0^{\rm IIA,cubic} = \frac{4}{3} t_1^3 + t_1^2 t_2 + 2t_1^2 t_3 + t_1 t_2 t_3 + t_1 t_3^2  
  }
and provides the classical (tree-level) contribution  in the weakly
coupled type IIA theory with $g_s^{\rm IIA}\ll 1$.

This type IIA model is dual to the heterotic $STU$-model, which has
been studied in  \cite{deWit:1995dmj}. Here $S$
denotes the (one-loop corrected) axio-dilaton and $(T,U)$ are the K\"ahler and complex
structure moduli of the $T^2$.
These three heterotic moduli  are related to complexified Kähler moduli on the
type IIA side as
\eq{
  t_1=iU\,,\quad   t_2=iS\,, \quad t_3=i(T-U)\,.
}
The K\"ahler cone is ${\rm Im}(t_i)>0$ which means that on the  heterotic
side we are in the chamber ${\rm Re}(T)>{\rm Re}(U)$.
The prepotential is given by
\eq{
     \label{prepothet}
    \mathcal{F}_0^{\rm het} =i\bigg( -\mathcal{F}_0^{\rm het,cubic} -
    \frac{1}{(2\pi)^3} &\sum_{k,l = 0}^{\infty} 2 c_1(kl) {\rm Li}_3(
    e^{-2\pi(kT + lU)}) \\[0.1cm] &- \frac{2}{(2\pi)^3} {\rm Li}_3
    (e^{-2\pi(T-U)}) + O\left(e^{-2\pi S}\right) \bigg) \,,
}
where the coefficients $c_1(n)$ are defined via
\eq{
    \frac{E_4 E_6}{\eta^{24}}(q) = \sum_{n = -1}^{\infty} c_1(n) = \frac{1}{q} - 240 - 141444 q + \ldots  \,.
}
The cubic part is 
\eq{
    \label{F0-het-cubic}
    \mathcal{F}_0^{\rm het,cubic} = S TU + \frac{1}{3} U^3 +  T^2 U \,.
  }
Note that only the first term is at tree-level in $g_s^{\rm het}$,
whereas the two $S$ independent ones and all remaining $S$
independent ones  from  \eqref{prepothet} make  the full one-loop
correction. The remaining unknown terms come from
heterotic $NS5$-brane instantons.
One can check that the type IIA and heterotic cubic terms directly
match reflecting the duality  $\mathcal{F}_0^{\rm IIA} =\mathcal{F}_0^{\rm het}$.

Moreover, exploiting this duality the Gopakumar-Vafa invariants in the
class
\eq{
  C=i(k T + l U)=(k+l) t_1 + k t_3
}
can be directly read-off from \eqref{prepothet}
\eq{
         \alpha_0^{(k+l,0,k)}=-2 c_1(kl) \qquad    {\rm for}\quad k,l\ge 0\, \ {\rm
           and}\ (k,l)=(1,-1)\,.
       }
Note that $l$ can also take negative values without violating the
Kähler cone condition.
Hence, we are in a situation where we happen to know the subset of  GV invariants for
$D2$-branes not wrapping the large base explicitly in terms
of a modular form. Note that this implies
that these GV invariants  grow like  $c_1(n)\sim \exp(4\pi\sqrt{n})$ for large $n$,
which is correlated with the emergent string limit that we are taking.

\subsubsection{Yukawa couplings from Schwinger integrals}

All these preliminary considerations were in some weakly coupled
limit. Now we analyse how the computation could be carried out
in the co-scaled  infinite distance limit $\tau_2\to \lambda \tau_2$,
$g^{\rm IIA}_s\to \sqrt\lambda g^{\rm IIA}_s$, where the 4D Planck scale remains
constant. In this limit, the light towers of states with a mass scale
below the species scale $\Lambda_{\rm sp}\sim M_{\rm pl}/\sqrt{\lambda}$
are strings arsing from $NS5$-branes
wrapping the $K3$-fiber and $D2$-$D0$ bound states
wrapping  2-cycles of the $K3$.
The contribution of these $D2$-$D0$ bound states to ${\cal F}_0$ is given by
the Schwinger integral \eqref{F0_Schwinger}.

To be concrete, let us discuss the zero point Yukawa-coupling $Y^{(0)}_{UUU}$.
To evaluate it, one first needs to determine all GV invariants
$\alpha_0^{(n_1,0,n_3)}$ for 1/2 BPS $D2$-branes wrapping
the 2-cycles in the $K3$-fiber. In general this is
a horrendous problem that only a ``CY-demon" can solve  by pure counting.
However, as just explained, we are in  the fortunate situation
that we know these GV invariants already from computations
in the weakly coupled regime. Hence, the ``CY-demon" will
eventually find the same numbers.
Therefore, for the  expression for the Yukawa coupling \eqref{derivF0} the demon will write
\eq{
  \label{yukawattta}
        Y_{UUU}=Y_{UUU}^{(0)} + 2\cdot 240 \sum_{l=1}^{\infty} l^3\, {q_2^l\over (1-q_2^l)} - \sum_{k,l=1}^{\infty} 2
    c_1(kl) \, l^3 {q_1^{k} q_2^l\over (1-q_1^k q_2^l)} - 2 \frac{q_1}{q_1-q_2} \,, 
}
where $q_1 = \exp(-2\pi T)$, $q_2 = \exp(-2\pi U)$ and we have used
$c_1(-1) = 1$ and  $c_1(0) = -240$. One realizes
that the second term on the right hand side  can be expressed via the Eisenstein series 
\eq{
    E_4(q) = 1 + 240 \sum_{n=1}^{\infty} \frac{n^3 q^n}{1-q^n} \,
  }
as $2(E_4(q_2)-1)$. According to \eqref{derivF0}, the diverging zero point Yukawa coupling is given by
\eq{
        Y^{(0)}_{UUU}&= 240 \sum_{l=1}^{\infty} l^3  -\sum_{k,l=1}^{\infty}  c_1(kl) \,
        l^3  - c_1(-1) \cdot (-1)^3\\[0.1cm]
         &= - 2\cdot 240 \sum_{l,d=1}^{\infty}  l^3  +2 \sum_{k,l,d=1}^{\infty}  c_1(kl)\, l^3
         + 2\sum_{d=1}^{\infty}  c_1(-1) \cdot (-1)^3\,,
       }
where, as for the example in section \ref{sec_regmod}, we introduced the sum
$\sum_{d=1}^{\infty} 1=\zeta(0)=-1/2$.
Following the same example, we can now regularize this expression by
introducing dummy variables
$p_1,p_2<1$ with $p_1<p_2$  and eventually take the limit $p_1,p_2\to
1$.
Thus, we get
\eq{
  Y^{(0)}_{UUU}\Big\vert_{\rm reg} := &\lim_{p_1, p_2\to 1}\left[
            - 2\cdot 240 \sum_{l,d=1}^{\infty}  l^3\, p_2^{ld}  +2 \sum_{k,l,d=1}^{\infty}
            c_1(kl) \,l^3\,  p_1^{kd} \,p_2^{ld}
            - 2\sum_{d=1}^{\infty} p_1^d \; p_2^{-d}\,\right]\\[0.1cm]
          = &\lim_{p_1, p_2\to 1}\left[
            2- 2E_4(p_2)  +2 \sum_{k,l=1}^{\infty}
            c_1(kl) l^3  {p_1^{k} \,p_2^{l}\over (1-p_1^{k} \,p_2^{l})}
            + 2  {p_1\over (p_1-p_2)}\right]\,.
        }
Note that the geometric series on the far right-hand side will only be convergent if $p_1<p_2$.
 Clearly, after these steps the regularized expression has the same form as
 the non zero-point Yukawa coupling in \eqref{yukawattta} but it is important
 to note that we are not taking the $p_1,p_2\to 0$ limit, where
 this expression nicely converges but the opposite $p_1,p_2\to 1$
 limit. In order to perform this limit and to isolate the singularity,
 we employ a useful  mathematical relation, the so-called 
 Harvey-Moore identity \cite{Harvey:1995fq}
\eq{
    \label{harvey-moore}
    - E_4(p_2) + \sum_{k,l,d = 1}^{\infty} c_1(kl)\, l^3\, p_1^{kd} p_2^{ld} + \frac{p_1}{p_1-p_2} = \frac{E_4(p_1) E_6(p_1)}{\eta^{24}(p_1)} \frac{E_4(p_2)}{j(p_2)-j(p_1)} \,,
  }
  where the right hand side is  a modular form of bi-degree $(-2,4)$.

 As for the pedagogical toy example in section \ref{sec_regmod}, we
 now write  $p_1=\exp(-2\pi/\Lambda_1)$,
$p_2=\exp(-2\pi/\Lambda_2)$ with $\Lambda_1<\Lambda_2$ and 
apply a modular $S$-transformation to the right-hand side of
\eqref{harvey-moore}.
In this way, we get
\eq{
  \label{regufinalyukhet}
  Y_{UUU}^{(0)} &=\lim_{\Lambda_1,\Lambda_2 \rightarrow \infty} \left(
  2-2{\Lambda_2^4\over \Lambda_1^2}
              \frac{E_4(i\Lambda_1)
                E_6(i\Lambda_1)}{\eta^{24}(i\Lambda_1)}
               \frac{E_4(i\Lambda_2)}{j(i\Lambda_2)-j(i\Lambda_1)}
               -{\rm Div}\right)\\
             &=\lim_{\Lambda_1,\Lambda_2 \rightarrow \infty} \left(
                  2 -2{\Lambda_2^4\over \Lambda_1^2}\Big(    1+
                   O(e^{-2\pi(\Lambda_2-\Lambda_1)})\Big)
                   -{\rm Div}\right)   \,.
 }                
 Here we realize the compelling  pattern that in the limit,
 the second term in \eqref{regufinalyukhet} contains only a
 diverging term and exponentially suppressed terms.
Hence, after minimally subtracting the divergent ${\Lambda_2^4/
  \Lambda_1^2}$ term and taking the limit we arrive at  the final result
\eq{
  Y_{UUU}^{(0)} = 2 \,,
  }
which is indeed the cubic term in the one-loop heterotic Yukawa
coupling \eqref{F0-het-cubic}. Recall that this corresponds
to a tree-level TIN in the type IIA
dual model.

Thus,  what we have effectively done here is the following:
since we do not have a ``CY demon", we used the knowledge of the GV invariants derived in the weakly coupled
region from the dual heterotic string (or as we will see later from
type IIA mirror symmetry)
to define the regularized value of the zero point Yukawa coupling,
arising from the Schwinger integral over $D2$-$D0$ bound states (in the strongly
coupled region), as 
\eq{
  \label{yukawak3}
    Y_{UUU}^{(0)} \vcentcolon=  -\lim_{U, T \rightarrow 0} \left[
    \big(i\,\del_U^3 \mathcal{F}_0^{\rm het} - \kappa_{UUU}\big) - {\rm Div} \right] \,,
  }
 where $\text{Div}$ denotes all terms that diverge in the  limit. 
 We note that  the cubic term  $\kappa_{UUU}$ can be obtained
  from the opposite limit
  \eq{
    \kappa_{UUU}=\lim_{U,T \rightarrow \infty} i\, \del_U^3 \mathcal{F}_0^{\rm het}\,.
  }
  The definition \eqref{yukawak3} means that one  gets
 precisely  $Y_{UUU}^{(0)} =\kappa_{UUU}$, if the $U,T \rightarrow 0$
 limit of
 $\del_U^3 \mathcal{F}_0^{\rm het}$ itself only
 gives divergent  and vanishing terms, i.e.\ no further constant terms. Thus, it is satisfactory that the proposed regularization  involves the
TIN $\kappa$ in an explicit way allowing us to formulate the previous simple
criterion.

 One could now use this definiton to evaluate the remaining
 three zero point Yukawa couplings $Y_{UUT}^{(0)}$,  $Y_{UTT}^{(0)}$
 and  $Y_{TTT}^{(0)}$.  Instead of doing this in detail, we just
 mention that $Y_{TTT}^{(0)}=0$ can be obtained
 from an analogous computation in the chamber ${\rm Re}(U)>{\rm
   Re}(T)$
 and an  analytic continuation of the ${\rm Li}_3(e^{-2\pi
   (T-U)})$-term in  \eqref{prepothet}.
  The couplings $Y_{UUT}^{(0)}$ and $Y_{UTT}^{(0)}$ will require 
 a different technique, as the Harvey-Moore formula cannot be applied directly.

To summarize, for an  emergent string limit we have successfully
invoked a regularization of the zero point Yukawa couplings
so that the Schwinger  integration over  the light
towers of $D2$-$D0$ bound states at strong coupling
reproduced a subset of the classical (tree-level) TINs.
Even though we are not yet taking the isotropic M-theory limit,
this computation serves as proof of principle of
how this regularization can be defined.
Of course we are not yet getting full emergence
as the term $\kappa_{STU}$ is at (heterotic $NS5$-brane) string tree-level
and not captured by the one-loop Schwinger integral.
The next step is to move forward to the isotropic
M-theory limit, where we expect the full weakly coupled type IIA tree-level
prepotential to be contained in the Schwinger one-loop integral
at strong coupling.

\section{Isotropic M-theory limit of CYs}

Now we want to generalize these computations to the ultimate case,
namely to isotropic M-theory limits
where the full homology lattice
contributes and where the  GV invariants generically scale exponentially as
$\exp(\gamma \beta)$.

Let us note that the example from the previous section
shares one essential
feature with the resolved conifold, namely that the
2-cycles wrapped by the light $D2$-branes are allowed to shrink to
zero size (in string units).
As is obvious from our construction, this feature was
crucial  in defining the regularization of \eqref{kappasum} via the limit
$t\to 0$ and employing the modular properties
of the Yukawa couplings to isolate the divergence and subtract it.
However, as already observed in the seminal work
\cite{Candelas:1990rm} and briefly reviewed in appendix \ref{app_a}, on the Quintic threefold, 
the
exponential degeneracy of the GV invariants is correlated
with the appearance of a divergence of the prepotential
not at zero  but at a finite value $t=t_c$ set by the location
of the conifold singularity. Note that this is not the minimal
value of ${\rm Im}(t)$ which is at the Landau-Ginzburg point.
Indeed, the Yukawa-couplings are regular at that point
so that it does not seem reasonable to regularize
our infinite sums by taking the limit to this point.

As shown in appendix \ref{app_b}, the conifold point 
is nothing else than the quantum corrected
singular point in the middle of the classical phase diagram
of a GLSM.
Thus,  in a certain sense we are indeed taking a limit towards
zero size, namely  towards vanishing FI parameter $\xi=0$.
Since  due to quantum effects this is not really possible, the most
natural thing to do is to take the limit towards its quantum corrected value.

\subsection{Emergence of TINs on CYs with $h_{11}=1$}

For concreteness, let us now consider CYs with one K\"ahler modulus
and  therefore no  fibration
structure. In this case, there is no obvious toroidal structure
present and one does not expect the Yukawa coupling  to  contain
any modular forms. As mentioned, the Yukawa couplings in the
large radius chart still have a divergence of the form
\eq{
   Y_{ttt}\sim {1\over (t-t_c) \log^2(t-t_c)} +\ldots \,,
   }
which is not at zero but at the
location of the conifold point $t=t_c$.
The same holds for the total list of 14 CYs with $h_{11}=1$
that we provide here for convenience (see e.g.\ \cite{Bonisch:2022mgw})
\eq{
  \label{listCYh11}
  &\mathbb P_{1^4,4,6}[12,2]_{2.302}\,,\
     \mathbb P_{1^3,2,5}[10]_{2.099}\,,\ 
   \mathbb P_{1^2,2^2,3^2}[6,6]_{1.874}\,,\   
   \mathbb P_{1^4,4}[8]_{1.695}\,,\ 
    \mathbb P_{1^3,2^2,3}[6,4]_{1.563}\,, \\[0.1cm] 
    &\mathbb P_{1^4,2}[6]_{1.421}\,, 
    \mathbb P_{1^5,3}[6,2]_{1.334}\,,\ 
    \mathbb P_{1^4,2^2}[4,4]_{1.255}\,,\    
    \mathbb P_{1^5}[5]_{1.208}\,,\ 
    \mathbb P_{1^5,2}[4,3]_{1.115}\,,\\[0.1cm] 
     &\mathbb P_{1^5}[4,2]_{1.029}\,, \ 
     \mathbb P_{1^6}[3,3]_{0.975}\,,\ 
     \mathbb P_{1^7}[3,2,2]_{0.891}\,, \ 
     \mathbb P_{1^8}[2,2,2,2]_{0.807}\,,
} 
where the lower index  is the value of ${\rm Im}(t_c)$ at the conifold singularity.
In fact, in \cite{Candelas:1990rm}  this
divergence was utilized to derive the asymptotic behavior
of the GV invariants
\eq{
    \label{GVgrowth}
    \alpha_0^{\beta} \sim  {e^{2\pi \beta \, {\rm Im}(t_c)} \over \beta^{3} \log(\beta)^{2}} \,\,,
  }
which is  briefly reviewed in appendix \ref{app_a}.  
Crucially, for defining a regularization of the zero point Yukawa
couplings, the divergence at the conifold singularity encodes
information about the exponential degeneracy of the
 GV invariants.
Therefore,  it does not make sense to define the regularization
of the zero point Yukawa coupling via a $t\to 0$ limit, but 
rather as
\eq{
  \label{yukawaviaconi}
        Y_{ttt}^{(0)} &:=-\lim_{t\to t_c} \left[\sum_{n=1}^\infty  \alpha_0^n \, n^3
          {q^n\over 1-q^n}-{\rm Div}\right]
               =-\lim_{t\to t_c} \left[\Big(\partial^3_t {\cal
                   F}_0\big|_{\rm weak}
                   - \kappa_{ttt}\Big)-{\rm Div}\right]
               }
with $q=\exp(2\pi i t)$ and having set $(2\pi i )^3/g_s^2=1$.
As usual,  before taking the limit to the conifold point, we minimally subtract the divergence
for  $(t-t_c)\to 0$.
To explain the logic, recall that the computation is now
done in the strong coupling M-theory limit, where we
integrate out the full tower of  1/2-BPS $D2$-$D0$ bound states.
Of course, only a ``CY demon" can do this in practise, but
we are lucky in that we happen to know the exact result
from the weakly coupled regime via non-renormalization theorems and
mirror symmetry. Thus, the result that the ``CY-demon" obtains
after carrying out  this infinite sum  should agree with the weakly coupled
result in the large complex structure regime $\partial^3_t {\cal F}_0\big|_{\rm weak}$ minus the triple
intersection number $\kappa_{ttt}$, that is not contained in the sum
$\sum_{n=1}^\infty  \alpha_0^n \, n^3 {q^n/(1-q^n)}$.

Now, we have arrived at an expression that is very similar
to the other regularized expressions \eqref{sum_Eisenstein} and \eqref{yukawak3} encountered so far,
where we were dealing with modular forms.
Instead of applying  a modular S-transformation, next we have 
to evaluate  $\partial^3_t {\cal F}_0\big|_{\rm weak}$ around  the conifold
point and disentangle the divergent, the constant and the suppressed
contributions in the $(t-t_c)\to 0$ limit.
We have succeeded in defining a working regularization,
if the following {\it regularization condition} holds:
\begin{quotation}
 \noindent
 {\it The $t \rightarrow t_c$ limit of
 $\del_t^3 \mathcal{F}_0\vert_{\rm weak}$  only
 gives divergent  and vanishing terms, i.e.\ no further constant terms. Then \eqref{yukawaviaconi} yields  precisely  $Y_{ttt}^{(0)} =\kappa_{ttt}$.
 }
\end{quotation}

\subsubsection{Example: Quintic}

Let us consider the Quintic, i.e.\ $\mathbb P_{4}[5]$, as a concrete example, where we can rely
on well known results.
The weakly coupled Yukawa coupling  $\partial^3_t {\cal F}_0\big|_{\rm
  weak}$  was already computed in the seminal paper
\cite{Candelas:1990rm} using mirror symmetry. Here we are just
collecting
a couple of useful relations obtained in \cite{Candelas:1990rm}
and also in \cite{Huang:2006hq,CaboBizet:2016uzv} and refer the reader to the original literature.
The mirror of the Quintic has one complex structure modulus,
which is given by the deformation of  the degree five hypersurface constraint
\eq{
  \sum_{i=1}^5  z_i^5 - (5\psi) \,z_1 z_2 z_3 z_5 z_5 =0
}
in $\mathbb P^4$. The conifold singularity is located at $\psi=1$.
A basis of local periods can be determined as solutions to the Picard-Fuchs
equation  and in the large complex structure regime
can be expressed as
\eq{
  \pi_{{\rm LCS},1}&=\omega_0(z)=\sum_{n=0}^\infty {(5n)!\, z^n\over
    (n!)^5} \,\\
  \pi_{{\rm LCS},2}&=\omega_0 \log z +z\, \sigma_1 (z) \\
   \pi_{{\rm LCS},3}&= \omega_0 \log^2 z +2   z\,  \sigma_1(z) \log z  
   +  z\,\sigma_2(z)\\
    \pi_{{\rm LCS},4}&= \omega_0 \log^3 z + 3
     z\, \sigma_1(z)  \log^2 z
   + 3 z\, \sigma_2(z) \log z + z\,\sigma_3(z)
}
with $z=1/(5\psi)^5$ and $|z|< 5^{-5}$. The
$\sigma_i(z)=\sum_{n=0}^\infty a_{i,n} z^n$ are infinite series 
in $z$, starting with a constant. Their precise form will
not be important for our purposes.
The next step is to map these periods to an integral symplectic basis,
which allows to read off the mirror map and the
prepotential via
\eq{
       \Pi_{\rm LCS}=\left(\begin{matrix} X_0 \\ X_1 \\ F_1 \\
           F_0 \end{matrix}\right) =\omega_0 \left(\begin{matrix}
         1\\ t\\ \partial_t {\cal F}_0 \\ 2 {\cal F}_0-t \partial_t {\cal F}_0 
           \end{matrix}\right)\, .
}
Here $t=iT$ denotes the complexified K\"ahler modulus of the Quintic
that enjoys an expansion
\eq{
  t={X_1\over X_0}={1\over 2\pi i}\log z + O(z)\,.
}  
Inverting this mirror map one can express the prepotential
around the large complex structure point as
\eq{
       {\cal F}_0={5\over 6} t^3 + {25 i \zeta(3)\over 2\pi^3}+ {\rm instantons}\,.
}

The radius of convergence of the large complex structure patch is
precisely $|\psi|=1$ so that the conifold singularity lies
on the boundary. Therefore, for determining the limit of the
Yukawa coupling \eqref{yukawaviaconi} this patch is not convenient and
one needs to know the periods in the conifold patch and how the two
patches are correctly glued together. For the quintic this problem
has been solved e.g.\ in \cite{CaboBizet:2016uzv} and here we just
recall their results  up to  the level of detail needed for our
purposes.

In the conifold patch, it is convenient to introduce the coordinate
$u=1-\psi^{-5}$ which vanishes at the conifold locus.
Then one can solve the Picard-Fuchs equation in this coordinate,
obtain a basic of periods and finally determine a transition
matrix to an integral  symplectic basis  $\Pi_{\rm C}$, which on the overlap
with  the LCS patch matches the former basis $\Pi_{\rm LCS}$.
This integral symplectic basis takes the form
\eq{
  X_0 &=-\rho_1(u)  {u  \over 2\pi i} \log u +\rho_2(u) \,\\
  X_1 &=\rho_3(u) \,\\
  F_1 &=\rho_4(u) \,\\
   F_0 &=u\, \rho_1(u) \,,
     }
where  the $\rho_i(u)=\sum_{n=0}^\infty b_{i,n} u^n$ are infinite series
in $u$.  Note that the logarithmic term in $X_0$ leads
to the monodromy $X_0\to  X_0 -F_0$
around the conifold.

Now we can read off the K\"ahler  $t$ in terms
of the conifold coordinate $u$, via
\eq{
        t={X_1\over X_0}={b_{4,0}\over b_{3,0}}
        \sum_{m=0}^\infty (u\log u)^m \, c_m(u) \,,
      }
where $c_m(u)$ denote infinite series in $u$ with $c_{0,0}=1$.
Using the concrete values $b_{4,0}=i 1.29357...$, $b_{3,0}=1.07073...$
we obtain in the $u\to 0$ limit $t_c=iT_c=i 1.20812...$.
Note that at leading order we get
\eq{
  \label{difftversu}
     t-t_c=t_c\, c_{1,0}\, u \log u + \ldots\,.
   }
Next we recall  that the Yukawa coupling can be expressed as \cite{Candelas:1990rm}
\eq{
  \label{yucandel}
              \partial^3_t {\cal F}_0\big|_{\rm weak}={1\over
                  \omega_0^2}
              \kappa_{\psi\psi\psi} {1\over (dt/d\psi)^3}
}
with
\eq{
  \kappa_{\psi\psi\psi} = \left( \frac{2\pi i}{5}\right)^3 {5\psi^2\over (1-\psi^5)} \,.
}
Using $u=1-\psi^{-5}$ and the general structure of the periods at the
conifold, one realizes that the building blocks in \eqref{yucandel}
can be expressed as
\eq{
      {dt\over du}&= \log u \left( {d_{-1}(u)\over \log u}
        +\sum_{n=0}^\infty (u\log u)^n \,d_n(u)\right) \\
      {1\over \omega_0^2}&= \sum_{n=0}^\infty    (u\log u)^n \,e_n(u)\\
      {\kappa_{\psi\psi\psi}\over (\partial_\psi u)^3}&= -{1\over 25}
      {1\over u (1-u)^3} \,,
    }
where the $d_n(u)$ and $e_n(u)$ are again infinite series in $u$.
Putting everything together and expanding for small $u$, we find
the general expansion
\eq{
  \partial^3_t {\cal F}_0\big|_{\rm weak}= {1\over u \log^3 u}
  \left( \sum_{n,k=0}^\infty    {(u\log u)^n\over \log^k u}  a_{n,k}(u)
\right)\,.
}
Consistent with \cite{Candelas:1990rm}, one sees that the leading divergence is
\eq{
       \partial^3_t {\cal F}_0\big|_{\rm weak}\sim {1\over u \log^3 u}
       +\ldots \sim   {1\over (t-t_c) \log^2 (t-t_c)} +\ldots \,,
     }
where we used \eqref{difftversu}.   
 However, for $n=0$ all the constant terms in $a_{0,k}(u)=a_{0,k,0}+O(u)$
 lead to singular terms
 \eq{
   \partial^3_t {\cal F}_0\big|_{\rm weak}=\sum_{k=0}^\infty
   {a_{0,k,0}\over u \log^{k+3} u} + {\rm Reg}\,,
 }
which according to our philosophy are minimally subtracted. 
With all the remaining  terms being  regular, the final and essential
question is whether they all go to zero or whether there remains
a constant contribution.
That this is not the case can be straightforwardly  seen by noting that the
cancellation of the $\log u$ term requires $n=k+3$, for which
 still a term $u^{k+2} a_{k+3,k}(u)$ remains that goes to zero for
 $u\to 0$. Hence there is no constant term and our proposed
 regularization \eqref{yukawaviaconi} indeed yields
 \eq{
     Y_{ttt}^{(0)} &=-\lim_{t\to t_c} \left[\Big(\partial^3_t {\cal
                   F}_0\big|_{\rm weak}
                   - \kappa_{ttt}\Big)-{\rm Div}\right]=\kappa_{ttt}=5\,.
               }
We expect the computation to be completely analogous for the list 
of 14 CYs with $h_{11} = 1$ presented in \eqref{listCYh11}.

Having a closer look at the computation one realizes that
the avoidance of a constant contribution in the limit $t\to t_c$ can be traced
back to the appearance of the $u\log u$ term in the period
$X_0$, i.e.\ that $\log u$ is always accompanied by
a linear factor $u$. This is of course a consequence
of the monodromy around the conifold point.

It could be that our findings are just  an artifact of the
too simple choice of CY manifolds with one K\"ahler modulus.
Therefore, next we generalize the computation to CYs with
two K\"ahler moduli. Moreover, we also analyze whether
the {\it regularization condition} is only true for approaching
conifold singularities or whether it also holds for
more general degeneration limits.
For $h_{11}=2$ it also happens that the codimension one degeneration loci
intersect in more than one point so that the question
arises which point to choose and also along which family
of path it should be approached in taking the limit.
To approach these questions, in the following sections we
consider the two CYs $\mathbb{P}_{1,1,1,6,9}[18]$ and
$\mathbb{P}_{1,1,2,2,6}[12]$.

\subsection{Emergence of TINs on  $\mathbb{P}_{1,1,1,6,9}[18]$}

We start with  the elliptically fibered  CY manifold
from
section \ref{sec:exellCY}, namely $\mathbb{P}_{1,1,1,6,9}[18]$
with Hodge numbers $(h_{11},h_{21})=(2,272)$.
The two complex structure moduli of the mirror dual CY
are parametrized by the deformations of the hypersurface constraint
\eq{
      p=z_1^{18}+z_2^{18}+z_3^{18}+z_4^{3}+z_5^{2}- 18 \psi\, z_1 z_2 z_3 z_4
      z_5 - 3 \phi\, z_1^6 z_2^6  z_3^6 \,.
    }
Introducing the combinations 
\eq{
  \ov x = -\frac{1}{2^2 3^8} {\phi\over \psi^6}\,,\qquad   \ov y = -{1\over
    \phi^3} \,,
}  
the   two conifold degeneration loci of the manifold  are
\eq{
  \label{locconia}
  \Delta_1 &= (1 − \ov x)^3 − \ov x^3 \ov y =0\\
   \Delta_2 &= 1 + \ov y =0\,.
}
The LCS point is at $\ov x=\ov y=0$ and one can solve the
corresponding Picard-Fuchs equations in the vicinity of this point,
determine an integral  symplectic bases   of periods $\Pi^{\rm LCS}_I$,
$I=1,\ldots,6$, such that the mirror map
is
\eq{
  t_1= {\Pi^{\rm LCS}_2\over \Pi^{\rm LCS}_1}= {1\over 2\pi i} \log(\ov x) +\ldots\,,\qquad
  t_2= {\Pi^{\rm LCS}_3\over \Pi^{\rm LCS}_1}={1\over 2\pi i} \log(\ov y) +\ldots  \,.
}
Here $t_1$ measures the size of the toroidal fiber
and $t_2$ the size of the curve $\mathbb P^1\subset \mathbb P^2$.
Inverting the mirror map, the prepotential comes out as
\eq{
          {\cal F}_0\vert_{\rm weak}={3\over 2} t_1^3 +{3\over 2}
          t_1^2 t_2 
         +{1\over 2} t_1 t^2_2 + {135 i \zeta(3)\over 4\pi^3} + {\rm instantons}\,.
        }
We are particularly interested in the Yukawa couplings, for which
the exact B-side Yukawas were provided in
\cite{Candelas:1993dm,Candelas:1994hw}, which
in terms of  $(\bar{x},\bar{y})$ read \cite{Hosono:1993qy}
\eq{
  \label{yukawaconiinterbars}
    \kappa_{\bar{x}\bar{x}\bar{x}} &= \frac{9 i}{8
      \pi^3}\frac{1}{\bar{x}^3 ((1-\bar{x})^3 - \bar{x}^3
      \bar{y})}\,,\quad \ \kappa_{\bar{x}\bar{x}\bar{y}} = \frac{i}{8 \pi^3}\frac{3(1-\bar{x})}{\bar{x}^2 \bar{y}((1-\bar{x})^3 - \bar{x}^3 \bar{y})} \,, \\
    \kappa_{\bar{x}\bar{y}\bar{y}} &= \frac{i}{8 \pi^3}\frac{(1-\bar{x})^2}{\bar{x} \bar{y}^2((1-\bar{x})^3 - \bar{x}^3 \bar{y})} \,, \quad \kappa_{\bar{y}\bar{y}\bar{y}} = \frac{i}{24 \pi^3}\frac{1-3\bar{x}+3\bar{x}^2}{\bar{y}^2(1+\bar{y}) ((1-\bar{x})^3 - \bar{x}^3 \bar{y})} \,.
}
Note that these expressions are valid throughout the entire complex
structure moduli space.

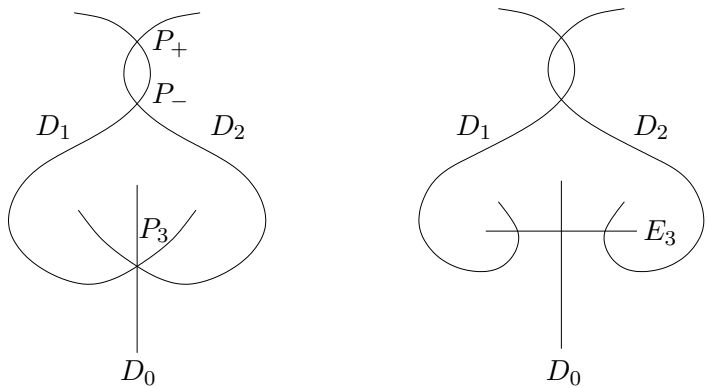
\begin{figure}[ht]
\centering
\includegraphics[width=0.6\textwidth]{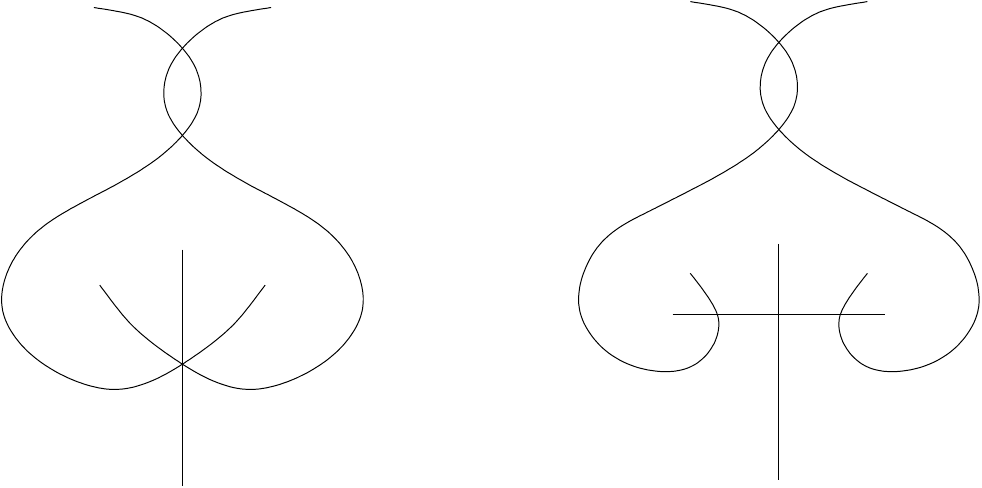}
\begin{picture}(0,0)
  \put(-225,-10){$D_0$}
  \put(-64,-10){$D_0$}
  \put(-257,83){$D_1$}
  \put(-191,83){$D_2$}
  \put(-98,83){$D_1$}
  \put(-31,83){$D_2$}
  \put(-27,43){$E_3$}
  \put(-213,115){$P_+$}
  \put(-213,95){$P_-$}
   \put(-218,44){$P_3$}
  \end{picture}
\vspace{0.5cm}
  \caption{Schematic view of the intersections of degeneration loci
    $D_1=\{\Delta_1=0\}$,  $D_2=\{\Delta_2=0\}$, $D_0$ and their resolution.}
\label{fig_Can}
\end{figure}

For regularizing the zero point Yukawa couplings \eqref{introeq} we now
want to take the limit to a point in the complex structure moduli space,
where the CY develops a singularity, which intuitively
encodes the asymptotic growth of the GV invariants in all directions
of the homology lattice $H_2(X,\mathbb Z)$.
The natural choice are the intersection points of the two conifold
loci \eqref{locconia}. According to \cite{Candelas:1994hw}, there exist three such points,
where the two points  $\ov x_0=(3\pm \sqrt{3} i)/6$ and $\ov y_0=-1$
are obvious and the third one at $\ov w_0=1/\ov x_0=0$, $\ov y_0=-1$
is actually a triple intersection with another orbifold singularity $D_0$
at $\psi=0$. This is  shown on the left  in figure \ref{fig_Can},
which is essentially taken from \cite{Candelas:1994hw}.

\vspace{0.0cm}

\noindent
To get a smooth moduli space with normal crossings of the
degeneration loci, one needs to perform a resolution of the triple
intersection. This is done via a blow-up, which introduces a divisor
$E_3$ which has normal crossings with the three degeneration loci
as shown on the right in figure \ref{fig_Can}.

\subsubsection{Emergence at the point $P_+$}

First, let us  consider the intersection point  $P_{+}$ with  $\ov x_0=(3+\sqrt{3} i)/6$
and $\ov y_0=-1$, where as shown in \cite{Alim:2012ss} it is convenient to
introduce the local coordinates
\eq{
       x_1=1-{\ov x\over \ov x_0}\,,\qquad\quad x_2=\left(1-{\ov x\over \ov
           x_0}\right)^{-1} \left(1-{\ov y\over \ov
           y_0}\right)\,
     }
 for which  the two  conifold degeneration loci become 
\eq{
  \label{conifoldlocs}
        \Delta_1={\textstyle {(1-i\sqrt{3})\over 2}} x_1 \big(1+O(x_i)\big)\,,\qquad\quad
        \Delta_2=x_1 x_2\,.
 }
 In appendix \ref{app_b}, we show that the intersection of the
 two conifold loci \eqref{locconia} corresponds
 to the quantum corrected singular point in the middle of the classical phase diagram
 (for the Fayet-Iliopoulos terms) of the  GLSM.
Note that $(x_1,x_2)$ should be regarded rather as
(complex) polar coordinates
around $P_{+}$, where $x_1$ is the radial direction and $x_2$ related
to  the angle, which is evident  from
\eq{
          {\ov y-\ov y_0\over \ov x-\ov x_0}={\ov y_0\over \ov x_0}
          x_2 = \tan\phi
        }
 and that $P_{+}$ is reached for $x_1=0$ independent of $x_2$.
 Like for radial coordinates this means that the map is not invertible
 at  $P_{+}$.

Then, our proposal is  that the TINs
$\kappa_{t_i t_j t_k}$ are  given by the regularization
\eq{
  \label{ex3modemerge}
  Y_{t_i t_j t_k}^{(0)}=-\lim \limits_{\substack{%
    t_1\to t_{1,c}\\
    t_2\to t_{2,c}}}
\left[\Big(\partial_{t_i}  \partial_{t_j} \partial_{t_k} {\cal
                   F}_0\big|_{\rm weak}
                   - \kappa_{t_i t_j t_k}\Big)-{\rm Div}\right]=\kappa_{t_i t_j t_k}\,.
               }
Therefore, to actually  evaluate the expression  \eqref{ex3modemerge}
we need to know the third derivatives of the prepotential 
in a vicinity of the intersection point $P_{+}$  of the two conifold loci.
For that purpose, we  first solve the Picard-Fuchs equations
in local coordinates around the point $(x_1,x_2)=(0,0)$ and then
determine an integral  symplectic basis which is required to be the continuation of the
integral symplectic basis around the LCS point $(\ov x,\ov y)=(0,0)$.
That means that on the overlap of two local charts around these
points, they have to coincide. To determine the numerical transition matrices with high enough precision, we have computed all period sets up to total order $40$ in the respective local variables $(x_1,x_2)$.

Similar computations have been performed  in \cite{Lee:2019wij,Alvarez-Garcia:2020pxd} and for our purposes we have been following this general recipe to arrive at the following results\footnote{We are grateful to Rafael \'Alvarez-Garc\'ia who helped us learning how to perform these computations.}, whose detailed derivation will be reported in \cite{unpublishedRafa}. Eventually, the periods around the point $(x_1,x_2)=(0,0)$ take the schematic form
\eq{
  \label{periodsconiinter}
  X_{\rm 0} &=x_1 P_0 \log(x_1) + Q_0 \,\\[0.1cm]
  X_{\rm 1} &=x_1 x_2 P_1 \log(x_1 x_2) + Q_1 \,\\[0.1cm]
   X_{\rm 2} &=x_1 x_2 P_2 \log(x_1 x_2) + x_1 \tilde P_2 \log(x_1)+Q_2 \,\\[0.1cm]
   F_{\rm 2} &=x_1 x_2 \hat P_2 \log(x_1 x_2) + \hat Q_2 \,\\[0.1cm]
   F_{\rm 1} &=x_1 \hat P_1 \log(x_1) + \hat Q_1 \,\\[0.1cm]
    F_{\rm 0} &=x_1 \hat P_0 \log(x_1) + \hat Q_0 \,,\\
  }
where the $P_i,Q_i,\hat P_i, \hat Q_i$ are polynomials in the
coordinates $x_1,x_2$.
The periods clearly feature the expected
logarithmic terms at the locations of the two conifold singularities
\eqref{conifoldlocs}.

\begin{figure}[ht]
\centering
\includegraphics[width=0.4\textwidth]{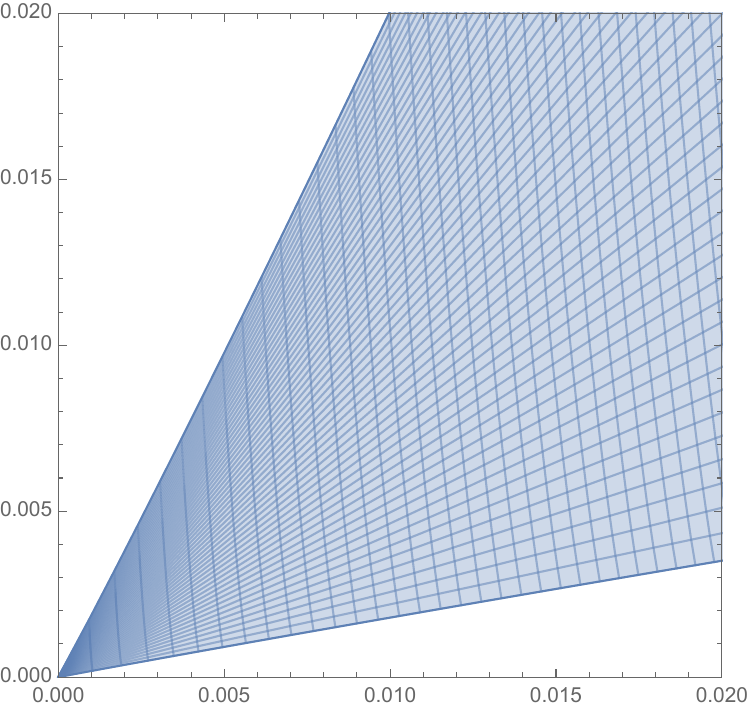}
\caption{Parametric plot of $({\rm Im}( t_1-t_1^{(0)}),{\rm
    Im}(t_2-t_2^{(0)}))$ for varying  $(x_1,x_2)$ with ${\rm Im}(x_i)=0$ and
    $0\le {\rm Re}(x_1)\le 0.1$,  $0.01\le {\rm Re}(x_2)\le 2.0$.}
\label{fig_1}
\end{figure}

Then, the two K\"ahler moduli enjoy the expansions
\eq{
  t_i= \sum_{n=0}^{\infty} \big(x_1 \log x_1\big)^n a_n^{(i)} +
  (x_1 x_2) \log(x_1 x_2) \sum_{n=0}^{\infty} \big(x_1 \log x_1\big)^n \, b_n^{(i)}  \qquad i = 1,2 \,,
}
where $a_n^{(i)},b_n^{(i)}$ are polynomials in $x_1,x_2$, where
however an $x_2^n$ term is always accompanied by at least one factor
of $x_1$. 
In the limit $x_1=0$,
the values of $t_i$ are determined by the constant terms in 
$a_0^{(i)}$ which give
\eq{
  \label{kaehlerpplus}
  t^{(0)}_1 \approx 0.137 + 0.991i\,,\qquad\qquad   t^{(0)}_2 \approx
  0.545 + 0.427i.
}
In figure \ref{fig_1} we show a parametric plot of  $({\rm Im}( t_1),{\rm
  Im}(t_2))$ around the singularity  at  $(t^{(0)}_1,t^{(0)}_2)$.
This shows that also in the K\"ahler moduli space the singularity can be approached
from various directions.
\vspace{0.0cm}
\noindent
Now we are ready to compute the Yukawa couplings \eqref{yukawaconiinterbars}.
For that purpose we first transform them
to the new coordinates $(x_1,x_2)$ and then to the
K\"ahler moduli  $(t_1,t_2)$ via 
\eq{
    \label{Kahler-Yukawa_coniinter}
      \partial_{t_i} \partial_{t_j} \partial_{t_k}
     \mathcal{F}_0 |_{\rm weak} = \frac{1}{\omega_0^2}
     \sum_{\alpha, \beta, \gamma} \frac{\partial 
       x_\alpha(t)}{\partial t_i} \frac{\partial  x_\beta(t)}{\partial t_j} 
     \frac{\partial  x_\gamma(t)}{\partial t_k} \kappa_{
       x_\alpha  x_\beta  x_\gamma}\,.
   }
Using \eqref{yukawaconiinterbars}, for the $\kappa_{x_\alpha  x_\beta  x_\gamma}$  we
find the behavior
\eq{
  \label{kappasings}
  \kappa_{x_1 x_1 x_1}&={1\over x_1} P_1(x_1,x_2) \,, \qquad
  \kappa_{x_1 x_1 x_2}= P_2(x_1,x_2)\,,\\
  \kappa_{x_1 x_2 x_2}&=x_1^2 P_3(x_1,x_2)\,,\qquad\;
   \kappa_{x_2 x_2 x_2}={x_1^2\over x_2} P_4(x_1,x_2) \,,
}
where the $P_i(x_1,x_2)$ are polynomials in $(x_1,x_2)$ having a
non-vanishing constant term.
   
For computing the partial derivatives $ \partial x_\alpha(t)/ \partial t_j$ one needs to the invert the mirror map,
which for the periods \eqref{periodsconiinter}
does not allow for an iterated procedure.
We circumvent this problem, by assuming that the $x_i$
are non-vanishing and applying the inverse function theorem
for holomorphic functions to determine the derivatives $\partial x_\alpha/ \partial t_j$
implicitly.
To do so, we need to invert the Jacobian of the mirror map
\eq{
    J_{\rm mir} (x_1,x_2)= 
    \begin{pmatrix}
    \frac{\partial t_1}{ \partial x_1} & \frac{ \partial t_1}{\partial
      x_2} \\
    \frac{\partial t_2}{\partial x_1} & \frac{\partial t_2}{\partial x_2} \\ 
    \end{pmatrix} \,,
}
allowing us to evaluate  \eqref{Kahler-Yukawa_coniinter}, where the right hand side 
is an expression depending on the complex structure coordinates
$(x_1,x_2)$, for which we eventually take the limit to $P_{+}$, i.e.\ $x_1\to 0$
for fixed $x_2$. More generally, we consider the limit $x_1\to 0$
along a generic path $x_2(x_1)=c + O(x_1)$ with the constant $c\ne 0$.

Similar to the computation for the Quintic, we have to keep track
of the leading singularities and the combinations of logarithmic
factors with linear terms in the $x_1$.
Doing this, we realize that we can express the derivatives
of the K\"ahler moduli as
\eq{
  \label{partialtss}
  {\partial  t_i\over \partial x_1}=(\log x_1)\,  {\cal G}^{(i1)}(x)\,,\qquad\quad
   {\partial  t_i\over \partial x_2}=(x_1 \log x_1) \,{\cal G}^{(i2)}(x)\
 }
with non-singular functions ${\cal G}^{(ij)}(x)$ having a general
expansion
\eq{
  \label{Gtypefunctions}
        {\cal G}^{(i,j)}(x_1,x_2)=\sum_{m,n,p=0}^\infty { (x_1 \log
          x_1)^m \over \log^p (x_1) } \left({ \log x_2\over  \log x_1}\right)^n a_{m,n,p}^{(ij)} \,,
}
where the  $a_{m,n,p}^{(ij)}$ are polynomials in $(x_1,x_2)$ having
a non-vanishing constant term. In \eqref{partialtss} most of the  $a_{m,n,p}^{(ij)}$
are actually vanishing but in the next step we have to invert
such expressions and then, like for the Quintic,  we generate the full
infinite series. Note that in the limit $x_1\to 0$ and fixed $x_2$
all terms in \eqref{Gtypefunctions} are controlled, i.e.
\eq{
  (x_1 \log(x_1))^m\to 0\,,\qquad \log^{-p}(x_1)\to 0 \,,\qquad
  \left({ \log x_2\over  \log x_1}\right)^n \to 0\,.
}

Now, inverting the Jacobian and using \eqref{kappasings} we  insert everything into
\eqref{Kahler-Yukawa_coniinter}, to realize that  we can write the
final results as a sum of four terms
\eq{
  \label{yuk-exp-11169}
  \partial_{t_i} \partial_{t_j} \partial_{t_k}
     \mathcal{F}_0 |_{\rm weak}  = &\frac{1}{x_1  \log^3(x_1)} \, {\cal G}_1^{(ijk)}(x) +
  \frac{1}{x_1  \log^3(x_1) } \, {\cal G}_2^{(ijk)}(x)\\[0.2cm]
  &+\frac{1}{\log^3(x_1)}\,  {\cal G}_3^{(ijk)}(x)+ \frac{1}{x_1
    \log^3(x_1)} \, {\cal G}_4^{(ijk)}(x)\,.
}
Here the ${\cal G}^{(ijk)}_n(x)$ are again functions of the type \eqref{Gtypefunctions}.
The first, the second  and the last term show the familiar conifold singularity.
The third term directly  vanishes  in the $x_1\to 0$ limit.
As for the Quintic, it is now evident that independent of the
``angle'' $x_2$, in the limit $x_1\to 0$ there are no constant terms
arising from the expression \eqref{yuk-exp-11169}.
There are only divergent terms, which we subtract and
terms that go to zero.
Hence, applying the regularization \eqref{ex3modemerge} one correctly obtains the
four TINs
\eq{
  Y_{t_1 t_1 t_1}^{(0)}=9\,,\qquad Y_{t_1 t_1 t_2}^{(0)}=3\,,\qquad Y_{t_1 t_2 t_2}^{(0)}=1\,,\qquad Y_{t_2 t_2 t_2}^{(0)}=0\,.
}
One might be worried about taking the limit
towards the degeneration point $P_{+}$
along a different family of paths and getting
a different result. This can indeed be the case and we will come
back to this issue in due course.
Moreover, we have obtained the desired results  by taking the limit towards the
intersection point  $P_+$, which after resolving the moduli space  is just one of the potential
degeneration points (see  \cite{Candelas:1994hw}).
For $P_-$ we expect that the computation goes through completely
analogous to $P_+$
However, for  handling the triple intersection point $P_{3}=D_1\cap D_2\cap D_0$ one has to perform the resolution
shown already in figure \ref{fig_Can}.
How this is  done locally is explained in \cite{Alim:2012ss}.
Since we have already identified a working intersection point,
we are not doing these  tedious computations for this CY
but will present an example   for $\mathbb{P}_{1,1,2,2,6}[12]$.

\subsection{Emergence of TINs on  $\mathbb{P}_{1,1,2,2,6}[12]$}
\label{sec_emergeCY2}

As a second example with two K\"ahler moduli let us consider the  CY manifold  $\mathbb{P}_{1,1,2,2,6}[12]$
with Hodge numbers $(h_{11},h_{21})=(2,128)$.
The GLSM data specifying the resolved  threefold are given in table
\ref{P11226data} (see \cite{Hosono:1993qy}).

\begin{table}[h]
    \begin{center}
    \begin{tabular}{|c|c c c c c c |c|} 
    \hline
    & $z_1$ & $z_2$ & $z_3$ & $z_4$ & $z_5$ & $z_6$ & $p$ \\ 
    \hline
    $l^{(1)}$ & 0 & 0 & 1 & 1 & 3 & 1 & -6 \\
    \hline 
    $l^{(2)}$ & 1 & 1 & 0 & 0 & 0 & -2 & 0 \\
    \hline
    \end{tabular}
    \caption{Data specifying $K3$-fibered CY}
    \label{P11226data}
    \end{center}
  \end{table}

  \noindent
 The threefold is a $K3$-fibration with base $\mathbb{P}^1$, where
   $z_1=0$  defines a divisor $L$ of topology $K3$ and  $z_3=0$
   a second divisor $H$, so that for this basis one
  gets the TINs  \cite{Candelas:1993dm} 
\eq{
    H^3 = 4 \,, \quad H^2 \cdot L = 2 \,, \quad H \cdot L^2 = 0 \,, \quad L^3 = 0 \,.
}
Expanding the K\"ahler form as $J=t_1\, \omega_H + t_2\, \omega_L$
one finds that $t_2$ measures the size of the $\mathbb P^1$ base
and that the K\"ahler cone is $t_1,t_2>0$.

As a  $K3$-fibration there exists a heterotic dual model so that the corresponding emergent
string limit is approached  for large radius of the base.
Then the two complex structure moduli of the mirror dual CY
are parametrized by the deformations of the hypersurface constraint
\eq{
      p=z_1^{12}+z_2^{12}+z_3^{6}+z_4^{6}+z_5^{2}- 12\psi\, z_1 z_2 z_3 z_4
      z_5 - 2\phi\, z_1^6 z_2^6\,,
}
where it is customary to  introduce the combinations 
\eq{
  \ov x = -\frac{1}{2^5 3^3} {\phi\over \psi^6}\,,\qquad   \ov y = {1\over \phi^2}\,.
}  
In terms of these, the manifold features  two degeneration loci at
\eq{
  \Delta_1 &= (1 − \ov x)^2 − \ov x^2 \ov y =0\,,\\
   \Delta_2 &= 1 − \ov y =0
}
where we notice that only the first one is a conifold singularity.
The LCS point is at $\ov x=\ov y=0$ and one can solve the
corresponding Picard-Fuchs equations in the vicinity of this point,
determine an integral symplectic basis   of periods $\Pi^{\rm LCS}_I$,
$I=1,\ldots,6$, such that the mirror map
is
\eq{
  t_1= {\Pi^{\rm LCS}_2\over \Pi^{\rm LCS}_1}= {1\over 2\pi i} \log(\ov x) +\ldots\,,\qquad
  t_2= {\Pi^{\rm LCS}_3\over \Pi^{\rm LCS}_1}={1\over 2\pi i} \log(\ov y) +\ldots  \,.
}
Inverting the mirror map, the prepotential enjoys the familiar
instanton expansion
\eq{
          {\cal F}_0\vert_{\rm weak}={2\over 3} t_1^3 +t_2 t_1^2
          + {63 i \zeta(3)\over 4\pi^3} + {\rm instantons}\,.
}
The exact B-side Yukawas were provided in
\cite{Candelas:1993dm,Candelas:1994hw}, which
in terms of $(\bar{x},\bar{y})$ read \cite{Hosono:1993qy}
\eq{
    \label{P11226-Yukawa-B-Bar}
    \kappa_{\bar{x}\bar{x}\bar{x}} &= \frac{2i}{\pi^3} \frac{1}{4\bar{x}^3 ((1-\bar{x})^2 - \bar{x}^2 \bar{y})}\,,\quad \phantom{aaaaaa}\kappa_{\bar{x}\bar{x}\bar{y}} = \frac{i}{2\pi^3} \frac{1-\bar{x}}{2\bar{x}^2 \bar{y}((1-\bar{x})^2 - \bar{x}^2 \bar{y})} \,, \\
    \kappa_{\bar{x}\bar{y}\bar{y}} &= \frac{i}{8\pi^3} \frac{2\bar{x}-1}{\bar{x} \bar{y}(1-\bar{y})((1-\bar{x})^2 - \bar{x}^2 \bar{y})} \,, \quad \kappa_{\bar{y}\bar{y}\bar{y}} = \frac{i}{32\pi^3} \frac{2(1-\bar{x}+\bar{y} - 3\bar{x}\bar{y})}{\bar{y}^2(1-\bar{y})^2 ((1-\bar{x})^2 - \bar{x}^2 \bar{y})} \,.
}

\subsubsection{Resolution of moduli space}
\label{sec_modulispace}

\begin{figure}[ht]
\centering
\includegraphics[width=0.6\textwidth]{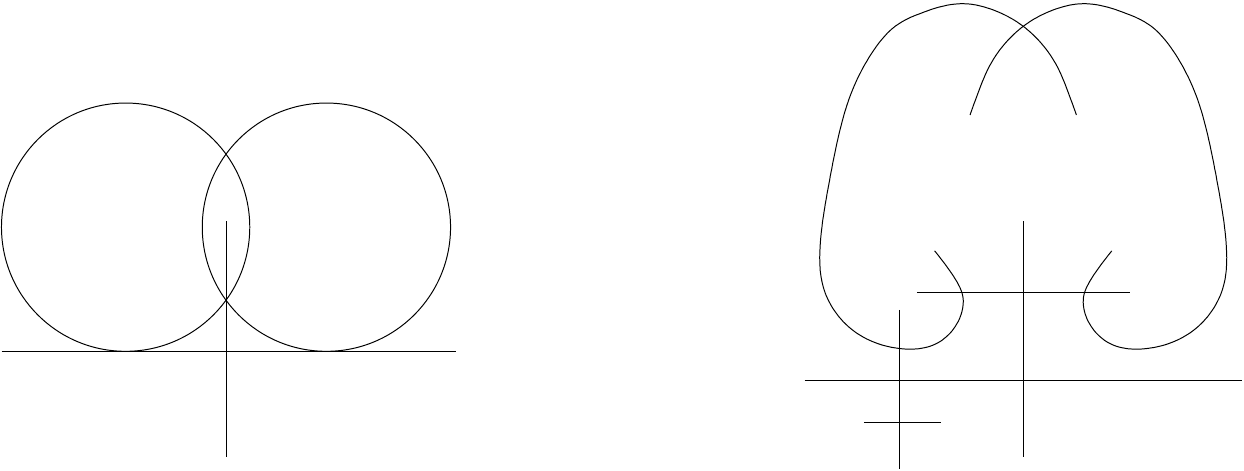}
\begin{picture}(0,0)
  \put(-225,-10){$D_0$}
  \put(-55,-10){$D_0$}
  \put(-83,-10){$E_1$}
   \put(-68,6){$E_2$}
  \put(-257,82){$D_1$}
  \put(-191,82){$D_2$}
  \put(-101,82){$D_1$}
  \put(-13,82){$D_2$}
  \put(-26,35){$E_3$}
  \put(-224,78){$P_1$}
  \put(-213,34){$P_2$}
  \put(-289,22){$D_\infty$}
   \put(-115,16){$D_\infty$}
  \end{picture}
\vspace{0.5cm}
  \caption{Schematic view of the intersections of degeneration loci
    $D_1=\{\Delta_1=0\}$,  $D_2=\{\Delta_2=0\}$, $D_0$, $D_\infty$  and their resolution.}
\label{fig_CanB}
\end{figure}
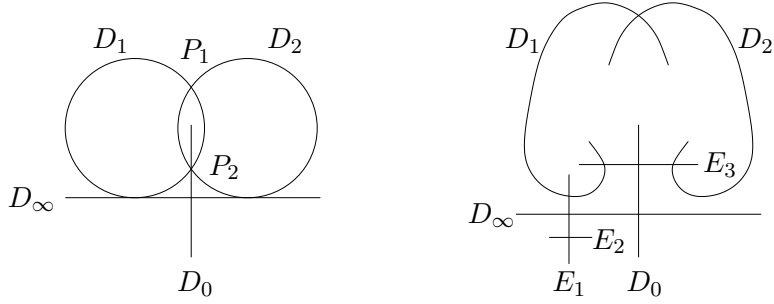

For the following analysis we need to collect some more information
about the resolution of the complex structure moduli space.
For the emerging string limit, we are interested in the
tangential intersection point between the conifold locus $D_1$
and the LCS locus $D_\infty$. As shown in figure \ref{fig_CanB},
its resolution involves two exceptional divisors with now
pairwise normal crossings.
The local coordinates  $(x_1,x_2)$ around the three intersections are \cite{Kachru:1995fv}
\eq{
  \label{crosstanginf}
   D_\infty\cap E_1: &\quad \bigg(  {\ov{x}^2 \ov{y} \over (1 -  \ov{x})^2},  1 - \ov{x}\bigg)\,,\\[0.1cm]
  D_1\cap E_1: &\quad\bigg( { (1 -  \ov{x})^2 - \ov{x}^2 \ov{y}
    \over (1 -  \ov{x})^2},  1 - \ov{x} \bigg)\,,\\[0.1cm]
   E_2\cap E_1: &\quad \bigg(  {\ov{x}^2 \ov{y} \over (1 -  \ov{x})},
   {(1 - \ov{x})^2\over \ov x^2\ov y}\bigg)\,.
   }
\vspace{0.0cm}

For the M-theory limit, the obvious intersection point $P_1$
of two degeneration loci $D_1$ and $D_2$ is at $(\ov x,\ov y)=(1/2,1)$,
but, according to \cite{Candelas:1993dm}, there exists a second  point $P_2$
at $(\ov w=\ov x^{-1},\ov y)=(0,1)$, where they intersect
also with a third  orbifold locus $D_0$ located  at $\psi=0$.
To get a smooth moduli space with normal crossings of the
degeneration loci, one needs to perform a resolution of this triple
intersection. This is done via a blow-up, which introduces a divisor
$E_3$ which has normal crossings with the three degeneration loci
as shown on the right in figure \ref{fig_CanB}.
How this is  done locally is explained in \cite{Alim:2012ss}.
To describe the resolved complex structure moduli space around $P_2$,
one needs three charts  where each chart provides the local coordinates $(x_1,x_2)$
for the normal intersection of the divisor $E_3$ with one of the  three degeneration loci
$D_1$, $D_2$ and $D_0$
\eq{
  P_{13}=D_1\cap E_3: &\quad\bigg(-{1\over \ov x} \big((1-\ov x)^2-\ov x^2
          \ov y\big),{1\over \ov x}\bigg)\,, \\
    P_{23}=  D_2\cap E_3: &\quad\bigg( \bar{x} (1-\ov{y}),  \frac{1}{\ov{x}} \bigg)\,,\\
         P_{03}=  D_0\cap E_3: &\quad \bigg( \frac{1}{\ov{x}(1-\ov{y})}, 1- \ov{y}\bigg)\,.
}
Note that in all three charts, the point $P_2$ with  $\ov w_0=1/\ov x_0=0$ and  $\ov y_0=1$
is reached for $x_2=0$ and $x_1$ can be regarded as an angular variable.

\subsubsection{Emergent heterotic string limit}
\label{subsec_emerghet}

For large $t_2$ this model has a weakly coupled heterotic dual on
$K3\times T^2$, where the heterotic 4d complexified dilaton and the K\"ahler
parameter of the $T^2$ factor are related to the two K\"ahler moduli
of the type IIA CY as
\eq{
  t_2=iS\,,\qquad   t_1=iT\,.
}  
Hence, $\ov y\to 0$ corresponds to the weak coupling limit on the
heterotic side. In the heterotic prepotential
\eq{
  \label{prepdualheto}
       {\cal F}_0^{\rm het} = i\bigg( -S T^2  -\left( {2\over 3} T^3 + O(e^{-2\pi T})
       \right) + O(e^{-2\pi S})\bigg) \,,
}
the first term is at tree-level, the second one the one-loop
correction
and the last one a sum over heterotic $NS5$-brane  instantons.
Now, let us analyze how (part of) the prepotential
is generated in  strongly coupled type IIA limits.

First, we take the emergent string limit on the type IIA side,
i.e.\ as in section \ref{sec_k3fibered} we co-scale the size of base like $t_2\to
\lambda t_2$ and the type IIA string coupling as
$g^{\rm IIA}_s\to \lambda^{1/2}g^{\rm IIA}_s$ so that the 4D Planck-scale
remains constant. This is the same limit as for the STU model
so that for the same reason not the full set of Yukawa-couplings
is emerging from integrating out wrapped $D2$-$D0$ bound states
at one loop.
In fact, emergence in this case means that the TIN $\kappa_{t_1 t_1 t_1}=4$ is given by the regularization of the
zero point Yukawa coupling
\eq{
  \label{ex2modemergeY}
      Y_{t_1 t_1 t_1}^{(0)}=-\lim_{t_1\to t_{1,c}} \lim_{t_2\to
        \infty} \left[\Big(\partial^3_{t_1} {\cal
                   F}_0\big|_{\rm weak}
                 - \kappa_{t_1 t_1 t_1}\Big)-{\rm Div}\right]\,.
 }
This is nothing else than the constant piece in the one-loop
correction to the Yukawa coupling \eqref{prepdualheto} on the dual heterotic side.
The first limit $t_2\to\infty$  guarantees that  before regularizing
the zero point Yukawa coupling via the limit  $\lim_{t_1\to t_{1,c}}$,
all $D2$-branes wrapping the large base $\mathbb P^1$ decouple.

This means that on the mirror dual side, we need to take the  limit
to the point $(\ov x,\ov y)=(1,0)$, which is the intersection
of the conifold locus and the LCS locus in $\ov y$.
Therefore, to actually  evaluate the expression  \eqref{ex2modemergeY}
we need to know $\partial^3_{t_1} {\cal F}_0\big|_{\rm weak}$
in a vicinity of this point in complex structure moduli space.
Since, the two loci intersect only tangentially one
needs to perform the resolution described in section \ref{sec_modulispace},
which introduces three local charts with coordinates
\eqref{crosstanginf}.

Let us consider the chart around $D_\infty\cap E_1$.
Again, one first needs to solve the Picard-Fuchs equations
in local coordinates  and then
to determine an integral symplectic basis which is required to be the continuation of the integral
symplectic basis around the LCS point $(\ov x,\ov y)=(0,0)$.
The resulting periods take the form
\eq{
  X_{\rm 0}/\omega_0 &= 1 \,\\
  X_{\rm 1} /\omega_0 &= P_1(x_1,x_2) \,\\
  X_{\rm 2}/\omega_0  &={1\over  2\pi i} \log(x_1 x_2^2) +{F_{\rm 2}/\omega_0\over
    2\pi i} \log(x_2) + P_2(x_1,x_2)\,\\
  F_{\rm 2}/\omega_0  &= \sqrt{x_2}P_3(x_1,x_2)  \,\\
  F_{\rm 1}/\omega_0  &=-{X_{\rm 1} /\omega_0\over \pi i}\log(x_1 x_2^2) + P_4(x_1,x_2) \, \\
   F_{\rm 0}/\omega_0 &= -\frac{1}{\pi i} \log(x_1 x_2^2) - \frac{F_2/\omega_0}{2\pi i} \log(x_1 x_2^2) + P_5(x_1,x_2) \,,
}
where each $P_i(x_1,x_2)$ is a power series in $(x_1,\sqrt{x_2})$. Indeed the periods $X_2,F_2$ feature the expected conifold behavior.
In the vicinity of $(x_1,x_2)=(0,0)$, the leading order behavior
of the K\"ahler  moduli is
\eq{
  \label{emergkaehler}
    t_1 &= {X_1\over \omega_0}= i + c_0 \sqrt{x_2} + \ldots \,, \\
    t_2 &= {X_2\over \omega_0} = {1\over  2\pi i} \log(x_1 x_2^2) + d_0 + d_1 \sqrt{x_2} \log(x_2) + \ldots \,,
}
with $c_i$ and $d_i$ series in $x_1$ and $\sqrt{x_2}$. Next one can determine the Jacobian and insert its inverse into the expression for the third derivative of the prepotential. Eventually, in the weak heterotic string coupling limit $x_1 \rightarrow 0$ this yields an expansion
\eq{
    \label{Y111_Het_A}
    \partial_{t_1}^3 {\cal F}_0={\, a_{-1}\over \sqrt{x_2}} +  a_0 \log\left( {x_2\over \Lambda_0}\right) +\sum_{n=1}^\infty \, a_n \log\left( {x_2 \over \Lambda_n}\right) x_2^{n/2} \,,
}
where the $a_n$ and $\Lambda_n$ are real with $a_{-1}\approx 4.826$, $a_{0}\approx 1.910$ and $\Lambda_0\approx 0.320$.
The first term diverges for the conifold limit $x_2 \to 0$ and
scales as
\eq{
  \label{emergentsingbev}
        Y_{t_1 t_1 t_1}^{(0)}\sim {1\over (t_1-i)} +\ldots \,.
      }
In appendix \ref{app_a}, we will relate this
to the asymptotic behavior of the GV invariants.
All the terms  on the right hand side of  \eqref{Y111_Het_A} go to
zero, while
the middle term also diverges though also containing  a constant term that is
however ambiguous\footnote{The constant piece in $\log(\epsilon A)$
  depends on the scale, i.e.\ it can be changed by rescaling
  $\epsilon\to \lambda \epsilon$ and then taking $\epsilon\to 0$ and
  minimally
  subtract the divergences.}. A very similar situation was also encountered
in the regularization of the Schwinger integral for the $R^4$-term in
8D
and the  one-loop topological free energy in 4D, where such an ambiguous
behavior was correlated with  the appearance of conformal symmetry \cite{Blumenhagen:2024ydy}. Hence, we consider such ambiguous  constants contained in $\log$-factors not as counter examples but as a sign that we are on the right track
and might just not yet perform the computation in the best suitable
chart. We indicate this by writing $ Y_{t_1 t_1 t_1}^{(0)}=4^*$.

To substantiate this idea we next  repeat
the above analysis in the chart around $D_1 \cap E_1$. The periods obey very similar relations and are given by
\eq{
  X_{\rm 0}/\omega_0 &= 1 \,\\
  X_{\rm 1}/\omega_0 &= R_1(x_1,x_2) \,\\
  X_{\rm 2}/\omega_0 &= \frac{1}{2\pi i} \log(x_2^2) + \frac{F_2/\omega_0}{2\pi i} \log(x_2) + R_2(x_1,x_2)\,\\
  F_2/\omega_0 &= \sqrt{x_2} \, R_3(x_1,x_2) \,\\
  F_1/\omega_0 &= - \frac{X_1/\omega_0}{\pi i} \log(x_2^2) + R_4(x_1,x_2) \,\\
  F_0/\omega_0 &= - \frac{1}{\pi i} \log(x_2^2) - \frac{F_2/\omega_0}{\pi i} \log(x_2) + R_5(x_1,x_2) \,. \\
}
Here each series $R_i(x_1,x_2)$ has the expansion
\eq{
    R_i(x_1,x_2) = \sum_{n=0}^{\infty} (x_1 \sqrt{x_2} \log(x_1))^n r^{(i)}_n (x_1,x_2) \,.
}
The $r^{(i)}_n$ are polynomials in $x_1$ and $\sqrt{x_2}$ and start with a constant. In the vicinity of $(x_1,x_2) = (0,0)$, the leading terms in the Kähler moduli are
\eq{
    \label{tmir_P_con}
    t_1 &= \frac{X_1}{\omega_0} = i + c_0 \sqrt{x_2} \, + c_1 x_1 \sqrt{x_2} \, \log(x_1) + \ldots \,, \\
    t_2 &= \frac{X_2}{\omega_0} = \frac{1}{2\pi i} \log(x_2^2) + d_0 + d_1 \sqrt{x_2} \, \log(x_2) + \ldots \,.
}
Also here $c_i$ and $d_i$ denote series in $x_1,\sqrt{x_2}$. Once again we want to evaluate $\partial_{t_1}^3 {\cal F}_0$ in the heterotic string weak coupling limit $\bar{y} \rightarrow 0$. In the present chart the coordinates satisfy the relation $\bar{x}^2 \bar{y} = x_2^2(1-x_1)$. In the limit $x_1 \rightarrow 1$ we reach the boundary of this chart, where we cannot trust the above periods any more. This suggests that we should rather take $x_2 \rightarrow 0$ (keeping $x_1$ small and constant), which according to \eqref{tmir_P_con} already ensures that $t_2 \rightarrow i \infty$ and $t_1 \rightarrow t_1^{(0)} = i$. The expansion of $\partial_{t_1}^3 {\cal F}_0$ for small $x_1,x_2$ yields 
\eq{
    \label{Y111_Het_B}
    \partial_{t_1}^3 {\cal F}_0 = \frac{f_0(x_1)}{x_1 \sqrt{x_2}\log^3(x_1)} + \frac{f_1(x_1) + f_2(x_1) \log(x_2)}{x_1 \log^3(x_1)} + \ldots \,,
}
where $f_i(x_1)$ are functions of $x_1$ and ellipses denote terms that
vanish for $x_2 \rightarrow 0$. We find the same leading singularity
in the variable $x_2$ as in \eqref{Y111_Het_A}. Moreover, there is a
series of $\log(x_2)$-divergent terms that are paired with constants
similar to the middle term in \eqref{Y111_Het_A}. However, unlike in
the previous case these constants all depend on $x_1$, but treating
$x_1$ as a constant we essentially get the same result
as for the chart $D_\infty\cap E_1$.

However, the absence of a bare, i.e.\ not $x_1$ dependent, constant in
\eqref{Y111_Het_B} makes it
possible to define a certain family of paths towards the singularity where $x_1$
scales as well. If one for example chooses $x_1 = z \
\sqrt{x_2}$ (with $z$ constant), no constant term remains after
minimally subtracting the divergences and taking the limit $x_2
\rightarrow 0$. Hence, the absence of a bare constant allowed us 
to find a family of paths that indeed yields
the desired value of the regularized  Yukawa coupling
$Y_{t_1 t_1 t_1}^{(0)}=4$.

\subsubsection{M-theory limit I}

So far all the limits we were considering for the two CYs with
two K\"ahler moduli were approaching conifold singularities.
For these the proposed {\it regularization condition} was satisfied.
In this respect the M-theory limit for $\mathbb P_{1,1,2,2,6}[12]$
is interesting, as it will involve taking the limit
to the intersection $\Delta_1\cap \Delta_2$ of the two degeneration
limits, where $\Delta_2$ is not a conifold locus, but a strong
coupling locus.
The obvious intersection point $P_1$ is at $(\ov x,\ov y)=(1/2,1)$,
but, as we have seen, there exists a second  point $P_2$
at $(\ov w=\ov x^{-1},\ov y)=(0,1)$, where they intersect
also with a third  orbifold locus at $\psi=0$.
The question is whether 
all four TINs do indeed  arise
from the regularization of the corresponding zero-point Yukawa couplings
$Y^{(0)}_{ijk}$.

To evaluate the limit towards $P_1$, we need to determine the integral symplectic basis of periods
around the intersection point $(\ov x,\ov y)=(1/2,1)$.
For that purpose it is useful to introduce the local coordinates
\eq{
    x_1=1-2\ov x\,,\qquad\quad   x_2={1-\ov y \over 1-2\ov x} \,,
}
where like for the coordinates of $\mathbb{P}_{1,1,1,6,9}[18]$ around
$P_{+}$, $x_2$ is an angular variable and the singularity is reached
by taking the limit $x_1 \rightarrow 0$ and $x_2 = c \neq 0$ kept
constant.
Then the two degeneration loci are at
\eq{
  \label{conidegslocs}
    \Delta_1=x_1 \big(1+O(x_i)\big)\,,\qquad\quad
    \Delta_2=x_1 x_2\,.
}
    
The periods take the form
\eq{
  X_{\rm 0}/\omega_0 &= 1 \,\\[0.1cm]
  X_{\rm 1}/\omega_0 &= -\frac{1}{2} X_2 + S_1(x) \,\\[0.1cm]
  X_{\rm 2}/\omega_0 &=\sqrt{x_1 x_2} \, S_2(x) \,\\[0.1cm]
  F_{\rm 2}/\omega_0 &= \frac{1}{2} F_1 - \frac{1}{2\pi i} X_2 \log(x_2) + \sqrt{x_1 x_2} \, S_3(x) \,\\[0.1cm]
  F_{\rm 1}/\omega_0 &= S_4(x) \\[0.1cm]
   F_{\rm 0}/\omega_0&= x_1 S_5(x) \,,
 }
where $(x)$ denotes dependence of $(x_1,x_2)$ and each of the series $S_i$ has the general expansion  
\eq{
    S_i(x) = \sum_{n=0}^{\infty} (x_1 \log(x_1))^n s_n^{(i)}(x)
}
and the $s_n^{(i)}$ are polynomials in $(x_1,x_2)$ starting with a constant. In the present case the two K\"ahler moduli share common terms and enjoy the expansion 
\eq{
  \label{kaehlerwerder}
  t_1 =  S_1(x_1,x_2) - \frac{\sqrt{x_1 x_2}}{2} \, S_2(x_1,x_2)
   \,,\qquad\quad  t_2 =  \sqrt{x_1 x_2} \, S_2(x_1,x_2) \,.
}
In particular, for both $S_1$ and $S_2$
each $x_2^n$ term is again always accompanied by at least one factor
of $x_1$. In the limit $x_1 \rightarrow 0$ one obtains
\eq{
  \label{pzerotvalues}
  t^{(0)}_1 \approx 1.334 i\,,\qquad\quad t^{(0)}_2 = 0\,.
}
Note that in contrast to the $\mathbb{P}_{1,1,1,6,9}[18]$ case,
here we see only the typical conifold behavior in $x_1$.
The other degeneration at $\Delta_1$ only leads to
power-law behavior. Close to $x_1=0$ the two
degeneration loci are mapped to $t_2=-2(t_1-t_1^{(0)})$ and
$t_2=0$ respectively. This means that close to $x_1=0$,  the
mirror dual of the LCS phase is the cone
\eq{
  C_{\rm LCS}=\Big\{ {\rm Im}(t_2)>0, {\rm
      Im}(t_1)>t_1^{(0)}-{1\over 2}{\rm Im}(t_2) \Big\}. 
}

\begin{figure}[ht]
\centering
\includegraphics[width=0.3\textwidth]{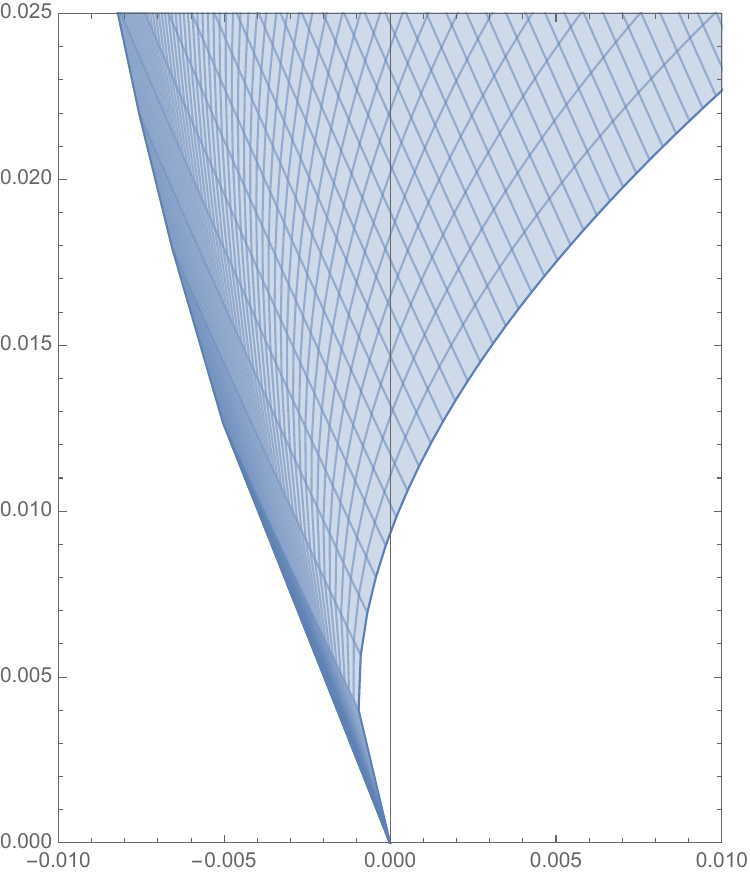}
\caption{Parametric plot of $({\rm Im}( t_1-t_1^{(0)}),{\rm
    Im}(t_2))$ for varying  $(x_1,x_2)$ with ${\rm Im}(x_i)=0$ and
    $0\le {\rm Re}(x_1)\le 0.1$,  $0.1\le {\rm Re}(x_2)\le 1$.}
\label{fig_2}
\end{figure}

Note that  compared to the $\mathbb P_{1,1,1,6,9}[18]$ there is an essential difference when approaching
the singular point in the K\"ahler moduli space.
From \eqref{kaehlerwerder} one can directly infer that for
$x_2$ constant and small $x_1$ one gets
\eq{
          {\Delta t_2\over \Delta t_1}=-2 +O(x_1) 
}
so that for any fixed value of $x_2$ one approaches the singularity
in a tangential manner.
This also evident from the parametric plot in
figure \ref{fig_2}, where
we observe that the shaded region is inside the cone
$C_{\rm LCS}$.
\vspace{0.2cm}

\noindent
Now we collect the remaining ingredients. Transforming the Yukawa couplings \eqref{P11226-Yukawa-B-Bar} to the new coordinates yields 
\eq{
  \label{kappasings2}
  \kappa_{x_1 x_1 x_1}&={x_2\over x^2_1 } P_1(x_1,x_2) \,, \qquad
  \kappa_{x_1 x_1 x_2}= \frac{1}{x_1} P_2(x_1,x_2)\,,\\
  \kappa_{x_1 x_2 x_2}&= \frac{1}{x_2} P_3(x_1,x_2)\,,\qquad\;
   \kappa_{x_2 x_2 x_2}={x_1\over x_2^2} P_4(x_1,x_2) \,.
}
Once again, the $P_i$ are polynomials in $(x_1,x_2)$ with non-vanishing constant term.

Next, we need to determine the leading behavior of the partial derivatives $\partial x_\alpha/ \partial t_j$. The derivatives of the mirror map scale as
\eq{
    \label{derivmir}
    \frac{\partial t_1}{\partial x_1} &= \log(x_1) \,
    \mathcal{G}_{1}(x) + \sqrt{x_2\over x_1}  \, \mathcal{H}_{1}(x)
    \,, \qquad  \frac{\partial t_1}{\partial x_2} = \sqrt{x_1\over x_2} \,\mathcal{G}_{2}(x)\,,\\[0.1cm]
    \frac{\partial t_2}{\partial x_1} &= \sqrt{x_2\over x_1} \,\mathcal{G}_{3}(x)\,, \qquad\phantom{aaaaaaaaaaaaia}
    \frac{\partial t_2}{\partial x_2} =  \sqrt{x_1\over x_2} \,\mathcal{G}_{4}(x)\,
}
with $\mathcal{G}_{i}$ and $\mathcal{H}$ all regular and finite in the
limit $x_1 \rightarrow 0$ with $x_2$ constant.
The inverse of the Jacobian matrix again yields the derivatives
$\partial x_\alpha/ \partial t_j$ as functions of $(x_1,x_2)$. Quite
remarkably, the leading-order term of the Jacobian determinant is not
a constant, as one would naively expect from \eqref{derivmir}. The
relations \eqref{kaehlerwerder} among the Kähler moduli lead to cancellations, so that the determinant scales as
\eq{
    \det J_{\rm mir} = \sqrt{x_1} \log (x_1) \, \mathcal{G}_{\rm det}(x) \,,
}
where $\mathcal{G}_{\rm det}$ is also regular in the limit we take.

Finally, we combine all the ingredients to obtain the Yukawa couplings
on the A-side. Evaluating \eqref{ex3modemerge} one obtains that for two
out of the four regularized zero point Yukawa couplings one indeed finds no constant
in the limit of $\partial_{t_i} \partial_{t_j} \partial_{t_k}
\mathcal{F}_0 |_{\rm weak} $ so that
\eq{
  \label{regyukcorr}
  Y_{t_1 t_1 t_1}^{(0)}=4\,,\qquad\quad Y_{t_1 t_1 t_2}^{(0)}=2\,.
}
However, for the other two, actually vanishing TINs
the regularization yields finite constant contributions so that
\eq{
  \label{regyuknotcorr}
  Y_{t_1 t_2 t_2}^{(0)}\approx 0.24\,,\qquad\quad Y_{t_2 t_2 t_2}^{(0)}\approx 0.36\,.
}
Hence, the proposed regularization method is only successful in half
of the cases.
For the two correctly regularized zero-point Yukawa couplings
\eqref{regyukcorr},
the  generic scaling of their  maximal divergence for $x_1\to 0$  is
\eq{
  \label{leadingdiva}
                       Y^{(0)}\sim {1\over x_1 \log^3(x_1)}+\ldots\sim
                       {1\over \big(t_1-t_1^{(0)}+{1\over 2} t_2\big) \log^2\big(t_1-t_1^{(0)}+{1\over 2} t_2\big)} +\ldots\,.
}
In appendix \ref{app_a} we investigate whether this behavior can be
directly related
to the asymptotic growth of the Gopakumar-Vafa invariants.

Looking at figure \ref{fig_2}, one might suspect that the mismatch
of \eqref{regyuknotcorr}
is related to the non-generic, tangential  limit that we are taking.
Then we should be able to improve
the results by finding more generic paths towards
the degeneration locus.
In the concrete case, such paths can be found by balancing
the two contributions for $\partial t_1/\partial x_1$ in
\eqref{derivmir}.
Approaching the singularity along paths with
\eq{
  \sqrt{x_1 x_2}=-z x_1 \log(x_1)\,,
}
we find  that indeed the family of paths is now more  generic
but the four resulting regularized zero point Yukawa couplings
 \eq{
  Y_{t_1 t_1 t_1}^{(0)}=4\,,\qquad Y_{t_1 t_1
    t_2}^{(0)}=2^{\star}\,,\qquad Y_{t_1 t_2 t_2}^{(0)}=0^{\star}\,,
  \qquad Y_{t_2 t_2 t_2}^{(0)}\approx 0.36+{0.15\over z}
}
are still not all correct.
Here the star again indicates that this is only up to conformal
scaling. As in the previous section \ref{subsec_emerghet}, the presence of bare
constants in \eqref{regyuknotcorr} only allows us to make some of them
ambiguous but not to get rid of them completely.

\subsubsection{M-theory limit II}

Next we consider taking the limit towards the singular point $P_2$ located at
$(\bar{x} = \infty\,, \bar{y} = 1)$, where the two degeneration loci
$\Delta_1$ and $\Delta_2$ intersect with the singular curve $D_0$ that
arises from an orbifold action on the moduli space
coordinates. As discussed in section \ref{sec_modulispace},
several blow-ups are required to achieve normal crossings
between those divisors so that,  after the resolution, there are three points of
intersection $\mathcal{P} \in \{P_{13},P_{23}, P_{03}\}$.

\begin{figure}[ht]
\centering
\includegraphics[width=0.5\textwidth]{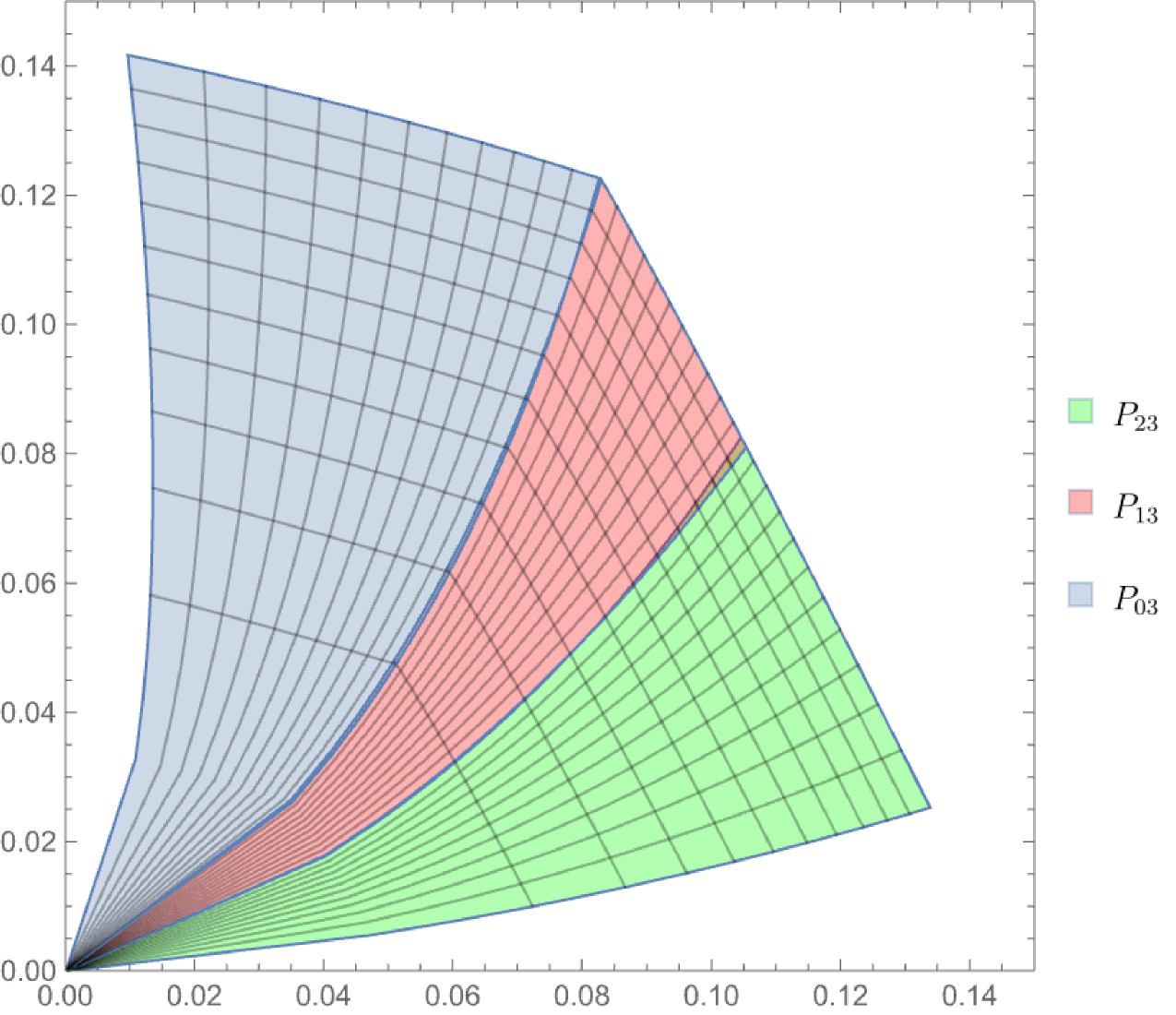}
\caption{Parametric plot of $({\rm Im}( t_1-t_1^{(0)}),{\rm
    Im}(t_2))$, evaluated in the patches around $\{P_{23},P_{13},P_{03}\}$. We vary the respective coordinates $(x_1,x_2)$ with ${\rm Im}(x_1)=0$, ${\rm Im}(x_2)=0$. For $P_{23}$ we have $0.1<{\rm Re}(x_1)<1$ and $0<{\rm Re}(x_2)<0.1$, for $P_{13}$ it's $0.01<{\rm Re}(x_1)<1$ and $0<{\rm Re}(x_2)<0.05$ and for $P_{03}$ we choose $0.1<{\rm Re}(x_1)<0.5$ and $0<{\rm Re}(x_2)<0.1$.}
\label{fig_5}
\end{figure}

We will now analyze the point $P_{03}$, which is the
intersection of $D_0$ with the  exceptional divisor $E_3$ separating the
curves parametrized by $\Delta_1$ and $\Delta_2$. To obtain the integral symplectic basis we solve the Picard-Fuchs  equations centered around each of the three above points and match the periods in suitable transition regions step by step. Around $(x_1,x_2) = (0,0)$, the periods take the form
\eq{
    X_0 &= (x_1 x_2)^{1/6} A_0 + \sqrt{x_1 x_2} \log(x_2) B_0 \\
    X_1 &= (x_1 x_2)^{1/6} A_1 + \sqrt{x_1 x_2} \log(x_2) B_1 \\
    X_2 &= x_1^{1/6} x_2^{1/2} C_2 \\
    F_0 &= (x_1 x_2)^{1/6} \hat{A}_0 + \sqrt{x_1 x_2} \log(x_2) \hat{B}_0 \\
    F_1 &= (x_1 x_2)^{1/6} \hat{A}_1 + \sqrt{x_1 x_2} \log(x_2) \hat{B}_1 \\
    F_2 &= (x_1 x_2)^{1/6} \hat{A}_2 + \sqrt{x_1 x_2} \log(x_2) \hat{B}_2 \,,
}
where $A_i,B_i,\hat{A}_i,\hat{B}_i$ denote polynomials in $(x_1^{1/3},x_2^{1/3})$ and each $x_1^{n/3}$ is paired at least with a factor of $x_2^{1/3}$. $C_2$ is a polynomial in $(x_1^{1/3},x_2)$ and contains bare $x_1$-terms. The Kähler moduli are given by the expansions
\eq{
    \label{mirmap_P01}
    t_1 &= \sum_{n=0}^{\infty} \big(x_1^{1/3} x_2^{1/3} \log(x_2)\big)^n
    a_n
       =t_1^{(0)}+x_2^{1/3} c_0 +x_1^{1/3} x_2^{1/3}\log(x_2) \, c_1+ \ldots\,, \\[0.1cm]
    t_2 &= x_2^{1/3} \sum_{n=0}^{\infty} \big(x_1^{1/3} x_2^{1/3}
    \log(x_2)\big)^n b_n
    =t_2^{(0)}+x_2^{1/3} d_0 + x_1^{1/3} x_2^{2/3}\log(x_2) \,d_1+ \ldots \,.
}
Here $a_n,b_n,c_n,d_n$ are also series in $(x_1^{1/3},x_2^{1/3})$. In
the limit $x_2 \rightarrow 0$ we obtain\footnote{Of course,  for
  $t_1^{(0)}$  we only obtain numerical values that are suspiciously
close (up to order $O(10^{-5})$) to the presented ones.}
\eq{
  \label{kaehlerpoint2}
  t_1^{(0)} = -{1\over 2} + {\sqrt{3}\over 2} i\,,\qquad\qquad t_2^{(0)} = 0\,.
}
This limiting value is common
for all three points from $\mathcal{P}$ when $x_2$ is
taken to zero, so that indeed we have three pathches around a common
intersection point $P_0$.
This is also shown in figure \ref{fig_5}.

To determine the Yukawa couplings we proceed as for the previous
cases. Transforming the B-side Yukawas to the coordinates of the patch
around $P_{03}$ yields
\eq{
  \label{kappasings3}
  \kappa_{x_1 x_1 x_1}&={x_2\over x_1} P_1(x_1,x_2) \,, \qquad
  \kappa_{x_1 x_1 x_2}= \frac{1}{x_1} P_2(x_1,x_2)\,,\\
  \kappa_{x_1 x_2 x_2}&= \frac{1}{x_2} P_3(x_1,x_2)\,,\qquad\;
   \kappa_{x_2 x_2 x_2}={x_1\over x_2^2} P_4(x_1,x_2) \,,
}
where each $P_i$ is a polynomial in $(x_1,x_2)$. We compute the partial derivatives $\partial x_\alpha/ \partial t_j$ again by using the derivatives of the mirror map, which are given by
\eq{
    \frac{\partial t_1}{\partial x_1} &= \frac{x_2^{1/3}}{x_1^{2/3}} \log(x_2) \, \mathcal{G}_1(x) \,, \\
    \frac{\partial t_1}{\partial x_2} &= \frac{1}{x_2^{2/3}} \mathcal{G}_2(x) \ + \frac{x_1^{1/3}}{x_2^{2/3}} \log(x_2) \, \mathcal{H}_2(x) + \frac{x_1^{2/3}}{x_2^{1/3}} \log^2(x_2) \, \mathcal{I}_2(x) + x_1 \log^3(x_2) \mathcal{J}_2(x) \,, \\
    \frac{\partial t_2}{\partial x_1} &= \frac{x_2^{1/3}}{x_1^{2/3}} \, \mathcal{G}_3(x) \,, \\
    \frac{\partial t_2}{\partial x_2} &= \frac{1}{x_2^{2/3}} \, \mathcal{G}_4(x) + \frac{x_1^{1/3}}{x_2^{1/3}} \log(x_2) \, \mathcal{H}_4(x) + x_1^{2/3} \log^2(x_2) \, \mathcal{I}_4(x) \,, \\
}
where each of the $\mathcal{G}_i,\mathcal{H}_i,\mathcal{I}_i,\mathcal{J}_i$ are regular and finite in the limit $x_2 \rightarrow 0$. The leading divergence of the Jacobian determinant scales like
\eq{
    \det J_{\rm mir} = \frac{\log(x_2)}{x_1^{2/3} x_2^{1/3}} \, \mathcal{G}_{\rm det} \,,
}
with $\mathcal{G}_{\rm det}$ also finite and non-divergent as before.

Finally, we combine all these results to evaluate the A-side zero
point Yukawa
couplings. We observe that their divergence 
only appears from the scaling  $X_0 \sim (x_1 x_2)^{1/6}$ of the
fundamental period.
Following the natural family of  paths with $x_1 = c \neq 0$ and $x_2 \rightarrow 0$, we find no
constant for two out of four regularized zero point Yukawa couplings, namely
\eq{
  Y_{t_1 t_1 t_1}^{(0)}=4\,,\qquad Y_{t_1 t_1 t_2}^{(0)}=2 \,.
}
For the remaining two Yukawa couplings we get no bare constants, but a series of $x_1$-dependent
terms that survive in the specified limit. Hence, we essentially
arrive that the same result
as for the M-theory limit I.

However, in contrast to the M-theory limit I, the absence of a bare constant and the experience from section \ref{subsec_emerghet} suggest that one can define another family of paths that works.
One such family is $x_1 = z \, x_2$ ($z$ constant), which ensures that both
Kähler moduli in \eqref{mirmap_P01} scale as $x_2^{1/3}$ at leading
order.
In this case all regularized zero point Yukawa couplings carry no constant
and really give the TINs
\eq{
    Y_{t_1 t_1 t_1}^{(0)}=4\,,\qquad Y_{t_1 t_1 t_2}^{(0)}=2 \,,\quad Y_{t_1 t_2 t_2}^{(0)}=0\,,\qquad Y_{t_2 t_2 t_2}^{(0)}=0 \,.
}
The  generic scaling of their  divergence for $x_2\to 0$  is
\eq{
  \label{singularbevb}
  Y^{(0)}\sim {1\over x_2^{1/3} \log^2(x_2)}\sim
  {1\over (t_i-t_i^{(0)}) \log^2(t_i-t_i^{(0)})}\,,
}
at least for three out of four zero point Yukawa couplings.

 The lesson we draw from this example is that the proposed
 regularization procedure is, maybe not surprisingly, sensitive to the
 actual co-dimension two degeneration point and the 
family of paths taken towards it.
 In contrast to the previous $\mathbb
P_{1,1,1,6,9}[18]$ CY,
in this example, we had to approach not the point $P_1$ but
$P_2$ to successfully obtain all four triple intersection
numbers via our proposed regularization method of the zero point Yukawa
couplings.
The art is to find a maximal codimension degeneration point
so that all regularized zero point Yukawa couplings do not contain
any bare constant. Then, it seems to be  possible to identify
a family of paths so that the proposal is satisfied
and the so defined regularization procedure gives the wanted
set of TINs.
The intuitive picture behind this is that by approaching the complex codimension 2
degeneration locus, the  limit has to be  sensitive to the asymptotic growth
of the GV invariants along all directions of the two-dimensional
homology lattice.
From the few examples studied so far, we could not yet identify a
clear pattern that allows us to decide beforehand what the best suited
patch around each degeneration point is.

\section{Conclusions}

The M-theoretic Emergence Proposal asserts that 
in the isotropic M-theory limit of type IIA, all interactions are generated
via quantum effects. This means that also the type IIA string tree-level
terms need to arise quantum mechanically.
This has been explicitly demonstrated for higher derivative
interactions, like $R^4$ and $F^4$ terms, in maximally respectively
half-maximally supersymmetric
theories, where they arose
at one-loop level from a real Schwinger integral, after
an appropriate regularization of the appearing UV divergences
had been implemented.

In this paper, we were trying to fill  a gap from  \cite{Blumenhagen:2023tev},
where a similar result was obtained for the  type IIA string tree-level
contribution to the $N=2$ prepotential in a 4D compactification
of type IIA on a non-compact CY threefold, namely
for the resolved conifold. There, the ultimate question remained,
how the classical TINs could
be obtained from a regularization of the GV-type one-loop Schwinger integral
for compact CYs. The problem is that even after regularizing
the Schwinger integral itself, eventually one has
to carry out a highly diverging infinite sum over the GV invariants.
Let us emphasize again  that the tree-level kinetic terms for the type IIA K\"ahler moduli have
their  origin in the 10D Einstein-Hilbert term so that we are really
dealing with the type of couplings mentioned in the original
Emergence Proposal \cite{Heidenreich:2017sim,Grimm:2018ohb,Heidenreich:2018kpg}.

After concretizing the question and taking some lessons from a toy model, we approached this problem in a step by step procedure. It turned out to be helpful to not directly
study the wanted M-theory limit but other infinite distance
limits where only part of the cubic terms in the prepotential
is expected to be generated at one-loop level.
CY manifolds featuring an elliptic or $K3$-
fibration structure were good candidates in this respect.
Generalizing the findings to the isotropic M-theory limit,
we managed to come up
with an admittedly still speculative but mathematically concrete proposal for how  the regularization
of this infinite sums can be performed, namely
\eq{
     Y^{(0)}_{t_i t_j t_k}=\frac{1}{2} \sum_{\beta \in H_2(X,\mathbb
       Z)}^{\infty} \alpha_0^{\beta}\, \beta_i \beta_j\beta_k \, \Big\vert_{\rm reg} =-\lim_{t_i \to t_{i,0}}
        \left[\Big(\partial_{t_i} \partial_{t_j}\partial_{t_k}{\cal
                   F}_0\big|_{\rm weak}
                 - \kappa_{t_i t_j t_k}\Big)-{\rm Div}\right]\,.\nonumber
}
We believe that it is satisfactory that the proposed regularization  involves the  TINs  in an explicit way.          
Recall that the appearance of the weakly coupled type IIA
prepotential is routed in the fact that, except for the triple
intersection
numbers to be subtracted explicitly, it implements the information about the
infinitely many GV invariants\footnote{A puzzling issue already mentioned in
\cite{Blumenhagen:2024lmo} is related to  the  Enriques CY, which is a free
quotient $X=(K3\times T^2)/\mathbb Z_2$ and known from \cite{Klemm:2005pd}
to have vanishing GV invariants $\alpha_0^\beta=0$. Clearly for this CY our
regularization method is void and it remains unclear how
the TINs can emerge from the now
vanishing zero-point Yukawa couplings.}. Effectively, 
we regularize the zero point Yukawa couplings via the
world-sheet instanton contributions to the Yukawas
and then take  a small radius limit towards a 
codimension $h_{11}$ degeneration locus.
For actually evaluating this expression we needed to determine
the  CY periods in a chart around the location $t_{i,0}$ of the
singularity, i.e.\ information that we generated by solving
the Picard-Fuchs equation and then continuously gluing
these solutions to  the periods in the well known large 
complex structure regime.

In this paper, we considered various degeneration loci of CYs with $h_{11}\le 2$. First, we were studying the Quintic CY, which
develops a conifold singularity at the degeneration locus (as do all 14 CYs with $h_{11} = 1$ from the list \eqref{listCYh11}).
We were also investigating in some detail the degenerations of two CYs with $h_{11}=2$, namely  the elliptic fibration $\mathbb P_{1,1,1,6,9}[18]$ and the $K3$ fibration  $\mathbb P_{1,1,2,2,6}[12]$. The first one featured the intersection of two conifold loci
and the second one the intersection of a
conifold and a strong coupling  locus.
While for the (intersecting) conifold loci the regularization quite
directly led to the wanted result $ Y^{(0)}_{t_i t_j t_k}=\kappa_{t_i
  t_j t_k}$, for the $K3$ fibration the analysis turned out to be more involved. More specifically, the result
of the above limit depended on the degeneration point and the local chart around it. We can summarize our findings in a {\it refined regularization condition}
\begin{quotation}
    \noindent
    {\it For each CY there exist a codimension $h_{11}$ degeneration
      point $t_{i,0}$ and a local chart around it so that 
      the  $t_i \rightarrow t_{i,0}$ limit of 
 $\partial_{t_i} \partial_{t_j}\partial_{t_k} \mathcal{F}_0\vert_{\rm
   weak}$ contains  no bare constants. Then, we expect to find
 a family of paths so that the limit gives only 
 divergent  and vanishing terms. Hence,
 the so defined regularization of the zero-point Yukawa couplings
 yields $Y_{t_i t_j t_k}^{(0)} =\kappa_{t_i t_j t_k}$. }
\end{quotation}
With only a few concrete examples studied in this paper, this should
be considered as a working hypothesis for more detailed studies in
the future, which should involve also CYs with more K\"ahler moduli and other
degeneration loci.

Let us emphasize again that in this paper  we were extending  the successful regularization
of UV-divergent real M-theoretic Schwinger integral from higher derivative terms
with extended supersymmetry 
to the 4D prepotential in theories with $N=2$ supersymmetry.
This  is to be considered  a pragmatic regularization method but lacks a truly finite
microscopic description of the M-theoretic one-loop amplitudes.
A proposal that goes in this direction 
has been put forward
in \cite{Hattab:2023moj,Hattab:2024thi,Hattab:2024chf} and it would be interesting to 
investigate any  relation between these two complementary approaches. Nevertheless, we think that the investigation reported in this paper has revealed
that the M-theoretic Emergence Proposal can be  a powerful guide to reveal
so far hidden structures  in the  relation
between the  asymptotic behavior of GV invariants and degenerations of CYs.
One could entertain the picture that the moduli spaces of
CY manifolds encode  so much information about quantum gravity that
their infinite distance degenerations are consistent with
the emergent string conjecture and that their finite distance
degenerations are consistent with the M-theoretic Emergence Proposal.

\vspace{0.0cm}

\paragraph{Acknowledgments:}
We thank Niccol\`o Cribiori for collaboration on earlier stages of this
work. Moreover, we are indebted to Rafael \'Alvarez-Garc\'ia for sharing
his insights into period computations with us.
We also acknowledge valuable discussions with  Murad Alim and
Antonia Paraskevopoulou.
The work of R.B. and A.G. is funded by the Deutsche Forschungsgemeinschaft (DFG, German Research Foundation) under Germany’s Excellence Strategy – EXC-2094 – 390783311.

\vspace{1cm}

\appendix

\section{Singularities and Gopakumar-Vafa invariants}
\label{app_a}

First we recall the  essence of the computation from \cite{Candelas:1990rm}.
Approaching the mirror dual of  conifold singularity in the Kähler
direction $t$, we know that the mirror map close to the singularity
has the behavior
\eq{
          t -t_c \sim u \log u \,.
}
Moreover, the Yukawa coupling $\kappa_{ttt}$ diverges like
\eq{
  \label{app_yuka}
                \kappa_{ttt}\sim {1\over u \log^3 u} \sim {1\over
                  (t-t_c) \log^2 (t-t_c)}   \,.             
              }
The question is how  the GV invariants $\alpha^n_0$ must scale with $n$
to reproduce this behavior from the relation
\eq{
         \kappa_{ttt}\sim \sum_{n=1}^\infty \alpha^n_0\, n^3\, e^{2\pi
           i n t}\,.
 }             
 Making an ansatz
 \eq{
         \alpha^n_0 \sim n^{\rho-3}\, (\log n)^{\sigma}\, e^{2\pi n
           {\rm Im}(t_c)} 
}
and approximating the sum over $n$ by an integral, one finds
\eq{
        \kappa_{ttt}&\sim \sum_{n}     n^\rho (\log n)^\sigma\,
        e^{-2\pi n {\rm
            Im}(t-t_c)}  \sim {\Gamma(1+\rho)\over (\Delta t)^{1+\rho}} (-\log
           \Delta t)^{\sigma}
}
with    $\Delta t=2\pi {\rm Im}(t-t_c)$. Comparing this to
\eqref{app_yuka} one concludes that the ansatz works for $\rho=0$ and
$\sigma=-2$. Hence, the GV invariants scale essentially exponentially
$\alpha^n_0\sim \exp(2\pi n {\rm Im}(t_c))$  for large $n$ and
we have verified that the divergence of the Yukawa couplings
at the conifold singularity contains the information about this behavior.

This argument was for a single K\"ahler modulus, but  should  be
generalized to more K\"ahler moduli. For the CY $\mathbb P_{1,1,1,6,9}[18]$
we have seen that there are two intersecting conifold loci so that
we expect that the GV invariants $\alpha_0^{(n_1,n_2)}$
counting BPS 2-cycles with volume $n_1 t_1 + n_2 t_2$
scale exponentially along both directions. Employing the
software package \texttt{CYTools} \cite{Demirtas:2022hqf, Demirtas:2023als},
we were determining the GV invariants up to total order $n_1+n_2\le
200$ with the lowest ones listed  in table \ref{table_a}.
\begin{table}[ht]
\centering
\includegraphics[width=1.0\textwidth]{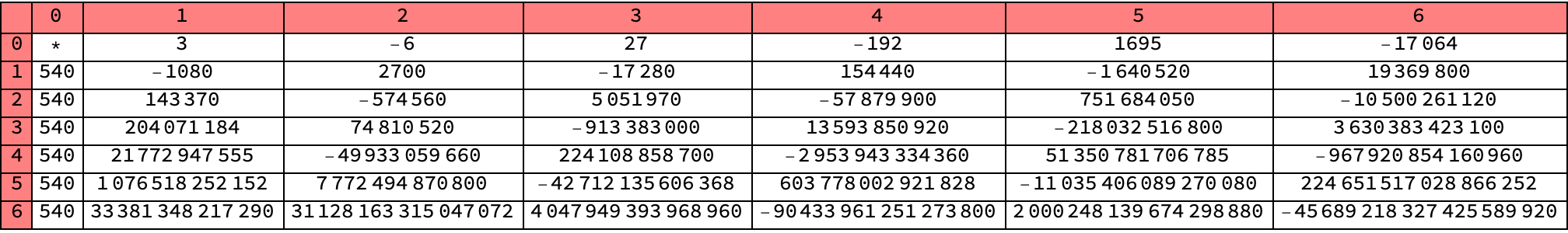}
\caption{Gopakumar-Vafa invariants of $\mathbb P_{1,1,1,6,9}[18]$.}
\label{table_a}
\end{table}
However, these data were not yet sufficient to fix the precise form
of the argument of the exponential function. Hence, also the
potential appearance of the values \eqref{kaehlerpplus}  of the K\"ahler moduli
$(t_1,t_2)$ at the point $P_+$ were not yet apparent.

For the other CY of interest, namely $\mathbb P_{1,1,2,2,6}[12]$ there
is an intersection of a conifold locus and another degeneration
locus at $t_2=0$.
In table \ref{table_b} we list the lowest GV invariants  $\alpha_0^{(n_1,n_2)}$,
which vanish for $n_2>n_1$ and satisfy the reflection property
$\alpha_0^{(n_1,n_2)}=\alpha_0^{(n_1,n_1-n_2)}$.
\begin{table}[ht]
\centering
\includegraphics[width=1.0\textwidth]{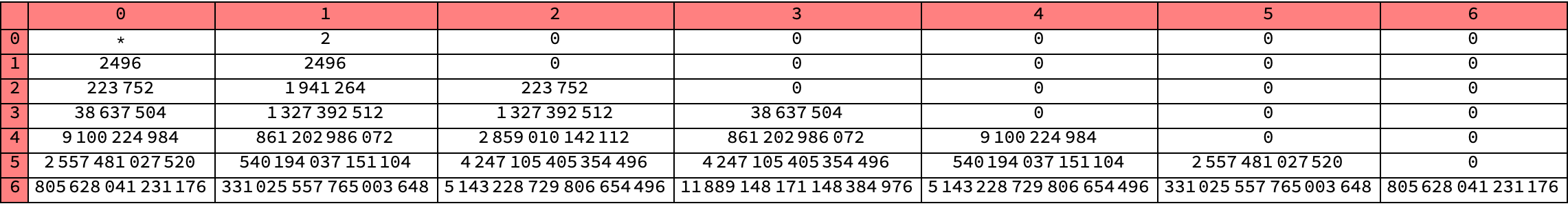}
\caption{Gopakumar-Vafa invariants of $\mathbb P_{1,1,2,2,6}[12]$.}
\label{table_b}
\end{table}

\noindent
In fact, we have computed all GV invariants up to total order
$n_1+n_2\le 200$.
Here the situation turns out to be a bit clearer than for
the previous CY.
Inspection reveals that for fixed $n_1$ the GV invariants
seem to  follow a Gaussian distribution with a maximum at
$n_2=[n_1/2]$
and width $\sigma^2\sim n_1$.
Looking more closer, after making an educated ansatz, we determine
the form of the GV invariants as
\eq{
  \label{ansatzGV2}
    \alpha_0^{(n_1,n_2)}\sim {e^{2\pi n_1 \mu }\,  {1\over n_1^{\rho+3}}\,
    \log^\sigma(n_1)} \exp\left( - 2\pi \lambda {(n_2-n_1/2)^2\over n_1}\right)\,
}
with  $\mu=\lambda=4/3$ and  $\sigma=0$.
We notice that $\mu=4/3$ precisely matches the value \eqref{pzerotvalues}
of ${\rm Im}(t_1)$ at the degeneration point $P_1$.
Moreover, for $n_2=0$ one can fit the GV data very well with $\rho=0$.
Just sticking to the first column in \ref{table_b}, the exponential
rate is $\mu-\lambda/4=1$, which happens to be the value \eqref{emergkaehler}
of ${\rm Im}(t_1)$ at the emergent string limit $D_\infty\cap D_1$.
Performing a similar computation as for the Quintic just for
this column we would get the singular behavior
\eq{
           Y_{t_1 t_1 t_1}\sim   {1\over t_1-i} +\ldots \,,
}
which indeed  matches \eqref{emergentsingbev} for the emergent string limit.

\begin{figure}[ht]
\centering
\includegraphics[width=0.3\textwidth]{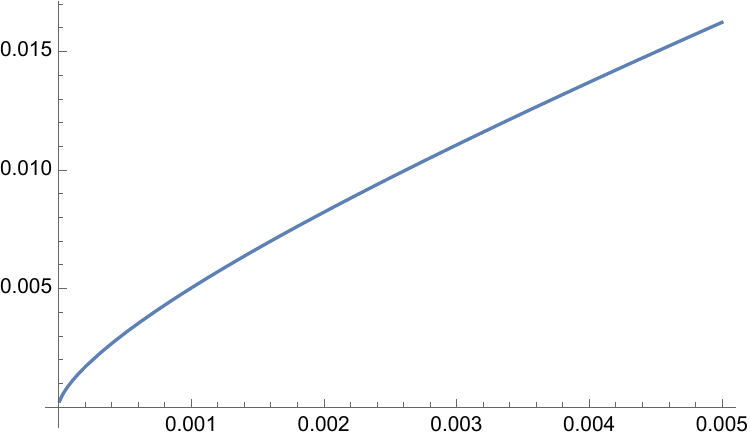}
\hspace{0.2cm}
\includegraphics[width=0.3\textwidth]{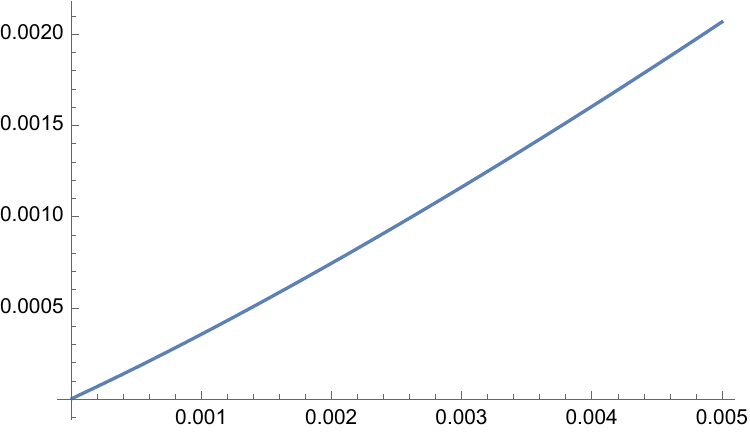}
\hspace{0.2cm}
\includegraphics[width=0.3\textwidth]{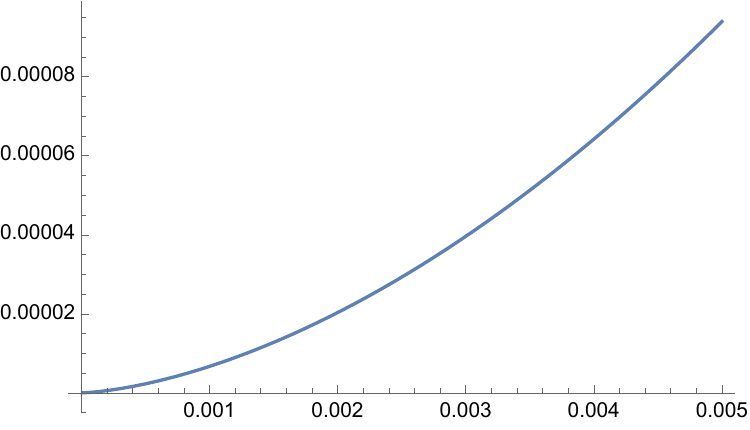}
\caption{Plots of $(\log^2(\Delta t)\, Y_{111})^{-M}$ as a function of
  $\Delta t$ for $\kappa=10$. Left $M=2/3$, middle $M=1$, right $M=3/2$. }
\label{fig_GVscale}
\end{figure}

Taking also the other columns into account, the data reveal
that $\rho$ is not constant but follows a plateau like
behavior that in the following we approximate via the function
\eq{
          \rho\approx{3\over 4}\tanh\left( \kappa {n_2(n_1-n_2)\over n_1^2} \right)\,,
        }
 where $\kappa$ is a not too small parameter to guarantee
 a broad plateau.
It is clear that the function can only depend on $n_2/n_1$
and must be invariant under the reflection $n_2\to n_1-n_2$.
Following the previous computation for the quintic, one could try
to estimate  the four Yukawa couplings via the double sum
\eq{
  \label{doublesum}
         Y_{t_1^l t_2^{3-l}}\sim \sum_{n_1,n_2=1}^\infty   n_1^l \,n_2^{3-l}\,
         \alpha_0^{(n_1,n_2)} \,e^{-2\pi (t_1 n_1 + t_2 n_2)} 
       }
with $l=0,\ldots,3$.       
Again, we can approximate the infinite sums via integrals over
continuous variables $(n_1,n_2)\to (y_1,y_2)$.
We did not succeed in analytically solving the appearing double integral.
However, after introducing polar coordinates $y_1=r \sin\phi$, $y_2=r
\cos\phi$
the integral over $r$ can be carried out analytically.
The result can then be numerically integrated over the angle
$0\le \phi\le \pi/4$ for various choices of $t_1$ and $t_2$
approaching zero. First, one realizes that \eqref{doublesum}
seems to diverge for $\Delta t=t_1-4/3+t_2/2$ to zero.
The precise functional dependence cannot be uniquely fixed
but we observe that the numerics is well consistent with a divergence
\eq{
  \label{mainsing}
         Y_{t_1^l t_2^{3-l}}\sim  {1\over \Delta t \log^2(\Delta t)}+\ldots\,.
}
This can be inferred from figure \ref{fig_GVscale}, which
should then be a straight line.
From  \eqref{kaehlerwerder} we find  at  leading order 
 \eq{
       \Delta t= t_1-t_1^{(0)}+t_2/2\sim x_1 \log(x_1) + \ldots
     }
so that we realize that \eqref{mainsing}  is consistent with 
the  leading divergence \eqref{leadingdiva}  found at the degeneration
point $P_1$. In the same manner it is consistent with  \eqref{singularbevb} for the point $P_2$
Having just this single example we cannot decide whether this is
really the correct procedure but  it  made it evident that the asymptotic
behavior of the GV invariants encodes information
about the degeneration loci, which is the underlying
structure of  our proposed regularization procedure
for the zero-point Yukawa couplings.

\section{Relation to singularities in Gauged Linear Sigma Models}
\label{app_b}

 In the proposed regularization method, it remains to better describe
 towards  which singular point  $t_{j,0}$ ($j=1,\ldots,h_{11}$)
 in K\"ahler moduli space the limit is taken.
 We observe that for all examples studied in this paper the CY
 manifolds admit a description in terms of Gauged Linear Sigma Models
 (GLSMs). The moduli space of these is parametrized
 by the complexified Fayet-Iliopoulos (FI) terms $\xi_j=\theta_j+i r_j$
 for the $h_{11}$ $U(1)$ gauge fields.
 The classical moduli space encounters  a singularity at the
 origin $\xi_j=0$, where all the different phases meet.
 However, due to quantum effects
 the FI-terms receive  one-loop corrections, which
 change the location of the singularity.
 In fact, one can use the algorithm of \cite{Morrison:1994fr} to determine
 the position of the quantum singularity.

Define $z_j=\exp(2\pi i \xi_j)$,  the quantum singular loci in the GLSM  are 
at the solutions to the  equations \cite{Morrison:1994fr}
\begin{equation}
\label{coni1}
\prod_{i=1}^M\big\langle\delta_i\big\rangle ^{Q_{i,j}}=z_j\,,\qquad j=1,\ldots,h_{11}\;
\end{equation}
and
\begin{equation}
\label{coni2}
\phantom{aa} \prod_{i\in I}\Big\langle\delta_i|_H\Big\rangle ^{Q_{i,j}}=z_j\,,\qquad
j=1,\ldots,{\rm rank}(H) \;,
\end{equation}
where $M$ denotes the number of $N=(2,2)$ chiral superfields and 
$\langle\delta_i\rangle$ the VEV of the operator
$\delta_i=\sum_{j=1}^{h_{11}} Q_{i,j}\,\sigma_j$. Here $\sigma_j$ is the
scalar part of the vector multiplet and  $Q_{i,j}$ the $G=U(1)^{h_{11}}$
charges of the chiral superfields.
Other components of singular loci are  \eqref{coni2} and are obtained from subgroups $H
\subset G$  such that the charges of the complementary set generate
(with positive coefficients) all of $\mathbb R^{h_{11}-k}$  where $k$ is the rank of $H$.
The product  in  \eqref{coni2} is over all chiral fields charged under
$H$ and the $\delta_i|_H$ are obtained
from $\delta_i$ by setting all scalar fields $\sigma$ related to the
complement of $H$ to zero.

The procedure is the same for all models, so we will demonstrate it in the case of the two parameter model $\mathbb{P}_{1,1,1,6,9}[18]$ with gauge group $G=U(1)\times U(1)$. The gauge charges are
\begin{equation}
  Q=\begin{pmatrix}
0 & 0 & 0 & 3 & 2 & 1 & -6 \\
1 & 1 & 1 & 0 & 0 & -3 & 0 
\end{pmatrix}\;.
\end{equation}
Inserting this into \eqref{coni1} and expressing the $\delta_i$ in terms of $\sigma_j$ gives
\begin{equation}
\begin{aligned}
z_1&=(-6s_1)^{-6} (3s_1)^3 (2s_1)^2 (s_1-3s_2)\\[0.1cm]
z_2&=s_2^3(s_1-3s_2)^{-3} \,,
\end{aligned}
\end{equation}
where for ease of notation $\langle \sigma_j\rangle=s_j$. These equations have solutions if 
\begin{equation}
  \label{quantshift1}
6^{12}z_1^3z_2-(1-3^3 2^4 z_1)^3=0\;.
\end{equation}
Moreover, the only subgroup of $G$ whose charges span $\mathbb{R}$ is
the first $U(1)$ so that $H$ is equal to the second $U(1)$.
Therefore \eqref{coni2} produces the additional relation
\begin{equation}
   \label{quantshift2}
z_2=s_2^3(-3s_2)^{-3}=-1/27\;.
\end{equation}
Hence, the two relations \eqref{quantshift1} and \eqref{quantshift2} specify the quantum shift of the classical
singularity at $\xi_1=\xi_2=0$.
Comparing this with the singularity in the mirror dual complex structure moduli 
\eqref{locconia}, we can read-off the identification
\eq{
         z_1={1\over 3^3\cdot 2^4} \,\ov x\,,\qquad\quad z_2={1\over 3^3} \,\ov y\,.
 }
This clearly shows that the intersection of the two conifold
singularities corresponds 
to  the quantum shifted singular point of the GLSM.
Note that in the regime of large K\"ahler moduli, one has an expansion \cite{Aspinwall:1994ay}
\eq{
      q_1=z_1 (1+O(z_1,z_2))\,,\qquad\qquad  q_2=z_2 (1+O(z_1,z_2))\,.
    }
The same analysis also works for the Quintic and $\mathbb P_{1,1,2,2,6}[12]$.

Therefore, for the case that we have a description of the CY in terms
of a GLSM  we have learned that in a certain sense we are indeed taking a limit towards
zero size, namely  towards vanishing FI parameters $\xi_i=0$.
Since  due to quantum effects this is not really possible, the most
natural thing to do is to take the limit towards their quantum corrected values.

\newpage

\vspace{3cm}

\bibliographystyle{utphys}
\bibliography{references} 

\end{document}